\documentclass[12pt]{article}%
\pdfoutput=1

\usepackage{amsmath}%
\usepackage{amsfonts}%
\usepackage{amssymb}%
\usepackage{graphicx}
\usepackage{color}
\usepackage{hyperref}
\usepackage[makeroom]{cancel}
\usepackage{authblk}

\usepackage{subcaption}

\newtheorem{lemma}{Lemma}

\begin{document}
\title{Caustics bounding entanglement wedges}
\author[1]{Marine De Clerck}
\author[1]{Charles Rabideau}
\author[1]{Niklas Tanger}
\vspace{1cm}
\affil[1]{\small{Theoretische Natuurkunde, Vrije Universiteit Brussel (VUB), and
International Solvay Institutes, Pleinlaan 2, B-1050 Brussels, Belgium.}}
\date{}

\maketitle

\begin{abstract}
We study the caustics on the boundaries of entanglement wedges in the context of holography in asymptotically AdS$_3$ spacetimes. These entanglement wedges play an important role in our understanding of the emergence of bulk locality. 
A procedure was proposed by Sanches and Weinberg for identifying boundary operators which are local in the bulk, which also applies to certain regions that lie beyond the reach of HRT surfaces by taking advantage of the lightsheets which bound entanglement wedges. 
We identify the caustics which terminate these lightsheets in conical deficit and BTZ black hole spacetimes and find that in some examples these caustics lead to a sharp corner in the entanglement wedge. 
The unexpected shape of these entanglement wedges leads, in those cases, to a breakdown of this procedure.
Many of the properties of the rich variety of caustics possible in higher dimensions remains to be explored which, as this work demonstrates, could lead to more unexpected features in the shapes of entanglement wedges. 

\end{abstract}

\tableofcontents

\section{Introduction}
Holography has provided insights into the emergence of locality in  quantum gravity. Early work on this topic includes the reconstruction of bulk operators using causal approaches \cite{BDHM98,HKLL06}. However, Ryu-Takayanagi (RT) surfaces \cite{RT06} reach outside of the region causally connected to a boundary subregion \cite{probes12,maximin12,dual_density_matrix12} and so it was appreciated that they must have some role to play in reconstructing the bulk from the boundary. Regions that are not crossed by RT surfaces, or their covariant generalisations due to Hubeny-Rangamani-Takayanagi (HRT) \cite{HRT07}, are known as an entanglement shadows \cite{entwinement14,shadows14}. 
However, the precise meaning of these regions is not fully understood. 

In recent years, an understanding of subregion-subregion duality in holography has lead to a new perspective on bulk locality. It has been understood that boundary locality leads to a well defined notion of local algebras of operators and that this local algebra can be associated to an appropriate algebra of bulk operators localised in the entanglement wedge 
associated with that boundary subregion \cite{HQECC14,JLMS15,EW_reconstruction16}. 

In this picture, HRT surfaces separate a bulk Cauchy slice into two parts, each reconstructible from complementary regions on the boundary. Because entanglement shadows can be contained in entanglement wedges, those bulk regions do not seem to be an obstruction to the reconstruction of the bulk from the boundary in this regard.

This analysis does not however have anything to say about whether the operators are localised at points in the bulk, it just restricts them to the entanglement wedge. 
Nonetheless, this machinery can be used to identify local operators \cite{stereoscopic16,KL17,SW,Locality_from_Modular17}. An operator that can be reconstructed independently in different boundary regions must lie in the intersection of the entanglement wedges of those regions. Therefore if a family of boundary regions can be found such that the intersection of their entanglement wedges includes only a single point in the bulk, then an operator that can be reconstructed in any of those regions must be localised at that point \cite{SW}. 
This work defined the localisable region as the set of bulk points that can be identified in this way.
Not all points in the bulk need to have this property and so the points that do not are known as non-localisable. 
The existence of local operators in semi-classical quantum gravity at non-localisable points cannot be established using this method. We will see that in some cases, causal reconstruction methods can be used to reconstruct operators in the non-localisable region. However, these causal methods only provide locality order by order in perturbation theory and lead to various confusions which were resolved using entanglment wedge reconstruction methods in \cite{HQECC14,EW_reconstruction16}. 
When the non-localisable region is behind a horizon it is not clear how to establish the existence of local operators.

The boundaries of entanglement wedges, which include the HRT surface, play an important role in the determination of the localisable region. However, the boundary of entanglement wedges also include the lightsheets emanating from the HRT surface towards the boundary and \cite{SW} proposed methods for localising points on these lightsheets. Clearly the non-localisable region is not in general the same as the entanglement shadows that were considered in \cite{entwinement14,shadows14}, but their interpretations have some similarities and we will see that they do coincide in some cases.

\subsection{Set-up}
Let us start by collecting the necessary notation. We will consider spacetimes, $M$, which are asymptotically AdS. Given an achronal subregion $R$ of the boundary, the set of points on the boundary spacetime for which every inextensible causal curve passing through the point also crosses $R$ is called the boundary domain of dependence of $R$, $D[R]$. The future/past domain of dependence of $R$ are denoted by $D^\pm[R]$. Let $J^\pm[S]$ denote the causal future/past of the subset $S$ of our spacetime. It was argued in \cite{BLR_CW12,HR_CW12,dual_density_matrix12} that a bulk field $\phi(x)$ can be reconstructed, to leading order in $1/N$, on a boundary subregion $D[R]$ whenever $x$ lies in the so-called causal wedge,
\begin{equation}
W_C(R) = \mathcal{J}^{+}[D[R]] \cap  \mathcal{J}^{-}[D[R]].
\end{equation}
of the subregion $R$. This bulk reconstruction method is known as causal wedge reconstruction.

We will denote the HRT surface anchored on a region $R$ by $\gamma_R$. This HRT surface can be taken to lie on a Cauchy slice of the bulk, $\Sigma_R$. It separates this Cauchy slice into $H_R$, the homology region connecting $R$ to $\gamma_R$, and $H'_R$.
$W_{\mathcal{E}}(R) \equiv D[H_R]$ is known as the entanglement wedge of $R$ and is the bulk region dual to $R$ in what is known as entanglement wedge reconstruction or subregion-subregion duality \cite{HQECC14,JLMS15,EW_reconstruction16}.

In \cite{SW}, a criterion was proposed for diagnosing whether a typically non-local boundary operator $\phi$ acting on a given code subspace $G$ (dual to an unknown bulk spacetime) corresponds to a local operator in the bulk. We will briefly summarise their proposal and recall the purpose of the localisable region, but refer to \cite{SW} for further details. The argument is based on the map $Q$ which associates the following set of boundary regions with $\phi$
\begin{equation}
    \mathcal{Q}(\phi)=\{R\in\mathcal{R}|\phi \text{\ is reconstructable in\ } R\}.
\end{equation}
This map defines equivalent classes $[\phi]$, where $\phi_1\sim\phi_2$ if and only if $\mathcal{Q}(\phi_1)=\mathcal{Q}(\phi_2)$. On these equivalent classes, one can associate the ordering $[\phi_1]\leq[\phi_2]$ if $\mathcal{Q}(\phi_1)\subseteq\mathcal{Q}(\phi_2)$. A set of operators $[\phi] \neq [1_G]$ with the property that for every operator $\phi'$ such that $[\phi] \leq [\phi']$ we also have $[\phi'] \in \{[\phi],[1_G]\}$, are called superficially local. This boundary characterisation of operators on the code subspace encodes the general intuition that the more local a bulk operator is, the more boundary regions it can be reconstructed on. In some sense, the superficially local operators in $[\phi]$ are as local in the bulk as it can be using the map $\mathcal{Q}$. However, not all superficially local operators are true local bulk operators\footnote{Note that the reverse is also true.}. 

The localisable region of the bulk (whose semi-classical Hilbert space is dual to the code subspace) is the subspace of bulk points for which superficial locality implies true bulk locality. A useful (bulk) criterion to determine if a bulk point $p$ belongs to the localisable region is proved as theorem III.1 in \cite{SW}, which will be a central tool in this work. It requires the existence of a subset of the collection of all boundary regions such that the intersection of their entanglement wedges contains only the point $p$, that is there is some family of boundary regions $\mathcal{R}_0$ such that
\begin{gather}
\bigcap_{R\in \mathcal{R}_0} W_{\mathcal{E}}(R) = \{p\}\,.
\label{eqn:SW_condition_localisability}
\end{gather}
In some asymptotically AdS spacetimes, such as pure AdS, the localisable region is the entire bulk. In that case, superficial locality coincides with locality. 

Points which are not in the localisable region are known as non-localisable. %
By taking the converse of \eqref{eqn:SW_condition_localisability}, a non-localisable point is one such that there exists another point $q$ so that  
\begin{align}
p \in W_{\mathcal{E}}(R)  \implies q \in W_{\mathcal{E}}(R)\, ,
\end{align}
for all $R \in R_0$.
Operators at these two points cannot be split into different entanglement wedges, so that the  argument for bulk micro-causality in \cite{HQECC14,JLMS15,EW_reconstruction16} does not apply.

\begin{figure}[th]
	\centering
		\includegraphics[width=.5\textwidth]{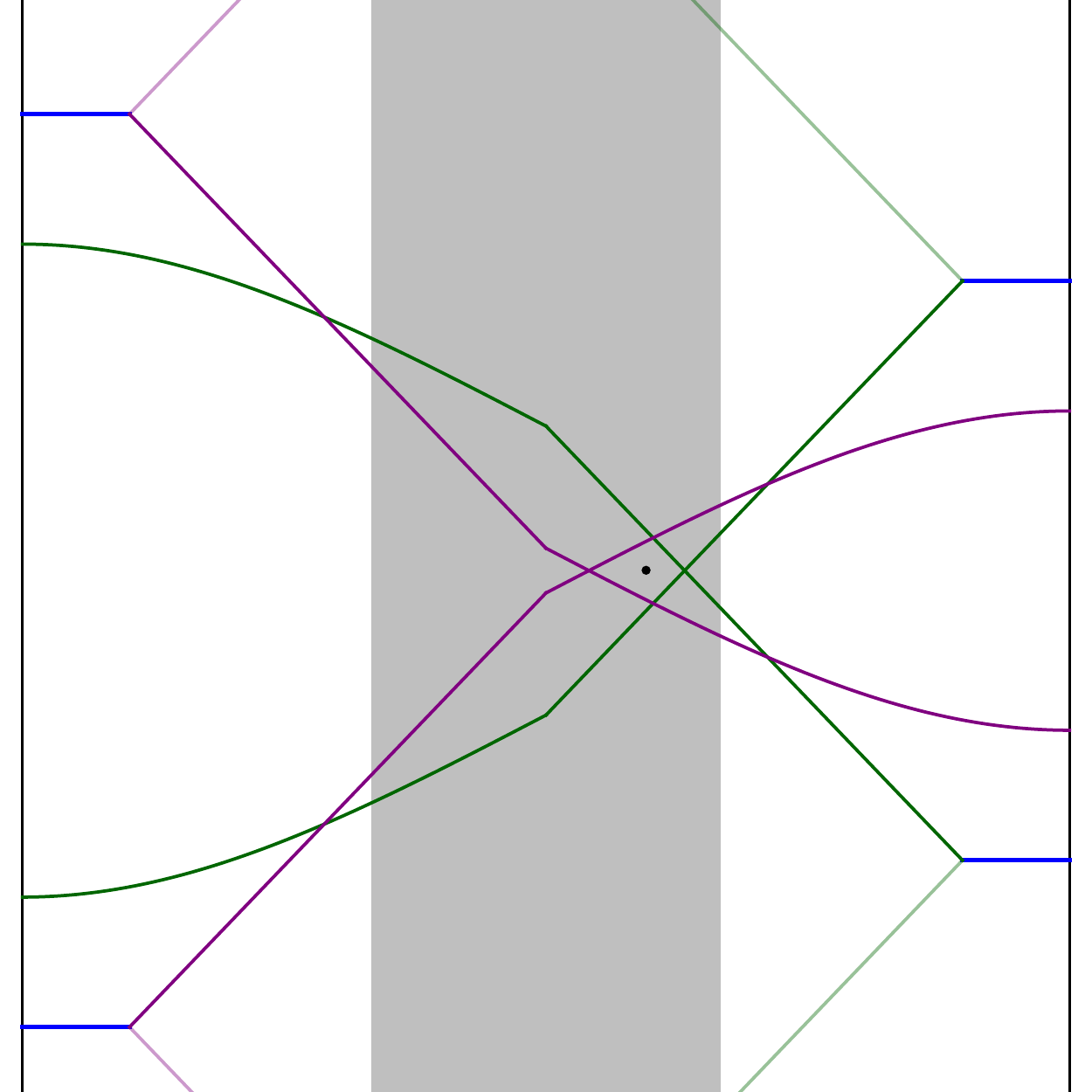}
		\vskip 2mm
    \caption{This figure is based on figure 3 of \cite{SW} demonstrating how to localise a point inside the entanglement shadow in the conical deficit spacetime. It depicts a conformal diagram, where light rays move at 45$^\circ$, of a $(r,t)$ slice of this spacetime with deficit angle $2 \pi/3$.
The entanglement shadow is shown in gray. Four HRT surfaces associated to large boundary regions are drawn in blue. 
The boundaries of the corresponding entanglement wedges are shown, in green and purple for HRT surfaces centered at $\theta = 0$ and $\theta = \pi$ respectively. This boundary is set by a light ray departing the HRT surface and reaching the defect at $r=0$. Behind the defect a caustic forms reaching the boundary at the other side of the cylinder. 
Provided that the corner between the light ray and the caustic is not too sharp, one can use such a set of boundary regions to localise a point in the entanglement shadow of a conical deficit.}
    \label{fig: assumption wedges}
\end{figure}

In a simple spacetime such as global AdS$_3$, where a Cauchy slice is completely probed by RT surfaces, finding a set of boundary regions with a single bulk point in the intersection of their entanglement wedges can be very simple. 
Namely, one can consider a bulk point as the intersection of two spatial geodesics. Because each geodesic is the set of points in the intersection of the two entanglement wedges bounded by that geodesic, the intersection of those four entanglement wedges contain only one bulk point. 
This is very similar to the intuition used to reconstruct bulk operators using invariance under modular flows proposed in \cite{KL17}. 
However, such an arguments only works for spacetimes entirely probed by RT surfaces, i.e. spacetimes without an entanglement shadow. 
It was argued in \cite{SW} that, even in the presence of an entanglement shadow, the entire bulk is in the localisable region when entanglement wedges probe the entire spacetime. 
This argument was based on an implicit assumption on the geometry of entanglement wedges, namely that the future and past boundaries of a cross-section of the entanglement wedge as depicted in figure \ref{fig: assumption wedges} are monotonic. 
In that case, it was argued that a set of entanglement wedges as shown in figure \ref{fig: assumption wedges}
would be sufficient to localise a point in the entanglement wedges of conical deficit spacetimes. 

In this work, we investigate this assumption by deriving the precise form of entanglement wedges and the caustics bounding them in asymptotically-AdS$_3$ geometries using the embedding space formalism. This will allow us to identify the non-localisable regions in some simple spacetimes by using the techniques proposed in \cite{SW}. As the lightsheets bounding entanglement wedges include caustics, an understanding of their shape is required to determine the extent of these non-localisable regions.

In section \ref{sec:cone}, we study entanglement wedges in the conical deficit geometry. We find an unexpected behaviour of the caustics bounding the entanglement wedges for deficit angles $\pi \leq \Delta \theta < 2\pi$ which implies a breakdown of the general analysis proposed in \cite{SW}. In this case, the non-localisable region coincides with the region not probed by HRT surfaces. For smaller deficit angles, the caustics have the behaviour anticipated by \cite{SW} and the whole spacetime is localisable.

In section \ref{sec:two-sided}, we derive the shape of some of entanglement wedges in the maximally extended two-sided BTZ black hole. Again, we will start by understanding the shape of the entanglement wedges and the caustics bounding them in this spacetime. HRT surfaces that stretch from one boundary to the other, which correspond to boundary regions including components in both boundaries, allowed \cite{SW} to localise points behind the horizon. Yet there is a region near the singularity that is not localisable. We prove a lemma demonstrating that entanglement shadows hidden behind event horizons lead to non-localisable regions. When given access to only one asymptotic region, as is the case for black holes formed by collapse, we find that there is a non-localisable region near the horizon which coincides with the entanglement shadow present in that case.

\section{Conical deficit}
\label{sec:cone}
The conical deficit spacetime is obtained by identifying the global angular coordinate $\theta$ of AdS$_3$ with $\theta+2 \pi \alpha$, with $0<\alpha<1$. 
Defining a new angular coordinate with the usual $2\pi$ periodicity,  while simultaneously rescaling the other global coordinates,
 one obtains the conical deficit metric,
\begin{equation}
\label{linecon}
    ds^2=-\left(r^2+\alpha^2 \right)dt^2+\frac{dr^2}{r^2+\alpha^2 }+r^2d\theta^2.
\end{equation} 

Determining the localisable region of a conical deficit spacetime necessitates understanding  the HRT surfaces and corresponding entanglement wedges associated to arbitrary boundary regions. These geometric constructs can be considered in the embedding space formalism, where AdS$_3$ is understood as a hyperboloid embedded in $\mathbb{R}^{2,2}$. We will use the convention that this space has signature $(-,-,+,+)$. The AdS hyperboloid is defined by $X^2=-L^2$, where $L$ is the AdS scale. This hyperboloid has a closed timelike curve which must be unravelled by taking its universal cover. We will work in units where $L=1$. Global coordinates on AdS$_3$ can be used to parametrise this hyperboloid as follows
\begin{align}
X^A_{global} (r,t,\theta) = \left( \sqrt{r^2+1} \cos t, \sqrt{r^2+1} \sin t,  r \sin \theta, r \cos \theta \right) \, .
\label{eqn:global_param}
\end{align}
The metric of AdS$_3$ in global coordinates is the one induced by the embedding of the hyperboloid into flat $\mathbb{R}^{2,2}$,
\begin{align}
ds^2 = dX \cdot dX = -\left(r^2+1\right) dt^2 + \frac{dr^2}{r^2+1} + r^2 d\theta^2 \,.
\end{align}

The identification of the angular coordinate required to obtain the conical deficit spacetime can be understood directly in the embedding space if one considers hyperpolar coordinates on $\mathbb{R}^{2,2}$ of the form
\begin{align}
\left( r_1 \cos \tau, r_1 \sin \tau, r_2 \sin \phi, r_2 \cos \phi   \right)\,.
\label{eqn:hyperpolar}
\end{align}
The action of $\phi \rightarrow \phi + 2\pi \alpha$ preserves the hyperboloid $X^2 = -1$. Once this identification is restricted to the hyperboloid, parametrised by \eqref{eqn:global_param}, it reproduces the usual identification of $\theta$ with $\theta+2 \pi \alpha$. The vector normal to 3-planes of constant $\phi_0$ is given by 
\begin{align}
P_{plane}(\phi_0) = \left(0 ,0 , \cos\phi_0 , - \sin \phi_0 \right)\, ,
\end{align}
and under the identification of the angular coordinate $\theta \sim \theta + 2 \pi \alpha$, the points $X$ satisfying $X \cdot P_{plane}(\phi_0)=0$ are being identified with those at $X \cdot P_{plane}(\phi_0+2 \pi \alpha)=0$.

A fundamental region of this identification can be covered by using global coordinates rescaled as
\begin{align}
\theta_{global} = \alpha \theta_{cone}\,, \qquad
t_{global} = \alpha t_{cone}\,, \qquad
 r_{global} = \frac{r_{cone}}{\alpha} \,,
\end{align}
with  $0<\theta<2 \pi$. This leads to a parametrisation of the AdS hyperboloid by 
\begin{align}
X^A_{cone} (r,t,\theta) = \frac{1}{\alpha} \left( \sqrt{r^2+\alpha^2} \cos \alpha t, \sqrt{r^2+\alpha^2} \sin \alpha t,  r \sin \alpha \theta, r \cos \alpha \theta \right) \, .
\label{eqn:cone_param}
\end{align}
The metric induced by the embedding into $\mathbb{R}^{2,2}$ reproduces the conical deficit metric \eqref{linecon}.

\subsection{HRT surfaces}
In AdS$_3$, HRT surfaces are given by spacelike geodesics.
The geodesics of a conical deficit spacetime can be obtained from the AdS$_3$ geodesics subject to the appropriate identifications. In the ambient $\mathbb{R}^{2,2}$ planes intersecting the hyperboloid $X^2=-1$ give the relevant geodesics \cite{integral_geometry15, causal_diamonds16}. 
Such a plane is spanned by the $\mathbb{R}^{2,2}$ vectors corresponding to a point on the geodesic, $X_0$, as well as the tangent at this point, $X_1$. Given a geodesic \begin{align}
\gamma^\mu(\lambda) = (r(\lambda),t(\lambda),\theta(\lambda)) \,,
\end{align}
the points on the geodesic are given by 
$X^A_{cone}(\gamma(\lambda))$, so that 
\begin{align}
 X_0 = X^A_{cone}(\gamma(\lambda_0)) \qquad \mathrm{and}\qquad X_1 \propto \partial_\lambda X^A_{cone}(\gamma(\lambda_0))
 \end{align}
  for a given reference point $\lambda_0$. For example, a geodesic with\footnote{This tangent vector, $\partial_\lambda \gamma^\mu(\lambda_0)$, is chosen for future convenience. Note that it has unit length and points in the $\partial_\theta$ direction for $\eta=0$. For $\eta\neq 0$ this HRT surface is anchored to an interval that is not centered at $\theta=0$.}
\begin{align}
  \gamma^\mu(\lambda_0) &= (r_0,\, 0,\, 0)\,, \\
  \partial_\lambda \gamma^\mu(\lambda_0) &= (-\eta \sqrt{r_0^2 + \alpha^2},\,  \frac{\eta}{\sqrt{r_0^2 + \alpha^2}} ,\,  \frac{1}{r_0}) \,.
  \end{align} 
 has 
\begin{align}
X_0 &= X^A_{cone} (r_0,0,0) = \frac{1}{\alpha} \left( \sqrt{r_0^2+\alpha^2}, 0,0, r_0  \right) \, , \\
X_1 &=  \partial_\lambda \gamma^\mu(\lambda_0) \partial_\mu X^A_{cone} (r_0,0,0)
   = \left(- \eta \frac{ r_0}{\alpha} ,  \eta, 1 ,-\eta \frac{ \sqrt{r_0^2+\alpha^2} }{\alpha} \right)
   \,.
   \label{eqn:hrt_cone}
\end{align}

This plane can also be described by the plane spanned by the vectors orthogonal to it, its normal space. 
A co-dimension 2 HRT surface always has a 2-dimensional normal plane. The 2-dimensional space normal to the HRT surface has one timelike and one spacelike direction, therefore the normal plane in $\mathbb{R}^{2,2}$ can be described either by a timelike and a spacelike unit vector $(S,T)$ such that $S^2 =1$, $T^2=-1$ and $S\cdot T=0$ or by a pair of null vectors $(N_1,N_2)$ such that $N_1^2=0$, $N_2^2=0$ and $N_1\cdot N_2 = -2$. These two descriptions are related by $N_1=T+S$ and $N_2 = T-S$. The HRT surface itself lives on a 2-plane spanned by  $X_0$ and $X_1$, with $X_0^2=-1$ and $X_1^2 = 1$.  so that $(X_0, X_1, S, T)$ form an orthonormal basis for $\mathbb{R}^{2,2}$. The intersection of this 2-plane with the AdS hyperboloid leads to a parametrisation of the HRT surface as
\begin{align}
Y(\xi) = \sec \xi X_0 + \tan \xi X_1 \qquad 
\mathrm{for}~-\frac\pi2< \xi< \frac\pi2 \,.
\end{align}
The HRT surface reaches the boundary for $\xi = \pm \frac{\pi}{2}$. These boundary points are described by the null rays $X_0 \pm X_1$ in the ambient $\mathbb{R}^{2,2}$. 
In terms of the parametrisation given in \eqref{eqn:hrt_cone}, the boundary points of HRT surfaces are described by the null rays in the direction of 
\begin{align}
X_0 \pm X_1 
= \left(
\frac{\sqrt{r_0^2 + \alpha^2} \mp \eta r_0}{\alpha} ,\pm \eta ,\pm 1, \frac{r_0 \mp \eta \sqrt{r_0^2 +	\alpha^2}}{\alpha}
\right)\,.
\end{align}
The conformal boundary of AdS$_3$ is given by null rays, $Z^2=0$ with $Z \sim \lambda Z$. In terms of the coordinates used to describe the conical deficit, this is
\begin{align}
Z^A_{cone} (t,\theta) \propto
\lim_{r\rightarrow \infty} \frac{\alpha}{r} X^A_{cone}(r,t,\theta) = \left(  
\cos \alpha t, \sin \alpha t, \sin \alpha \theta, \cos \alpha \theta
\right) \,.
\end{align}

Comparing these two expressions gives the endpoints of the boundary interval to which our HRT surface is attached
\begin{align}
\theta_\pm = \pm \frac{1}{\alpha} \arctan \frac{\alpha}{r_0 \mp \eta \sqrt{r_0^2 + \alpha^2}}
\,,\qquad 
t_\pm = \pm \frac1\alpha \arctan \frac{\eta \alpha}{\sqrt{r_0^2 + \alpha^2} \mp \eta r_0 }
\,.
\end{align}

\begin{figure}[th]
	\centering
	\begin{subfigure}[b]{0.45\textwidth}
		\includegraphics[width=.9\textwidth]{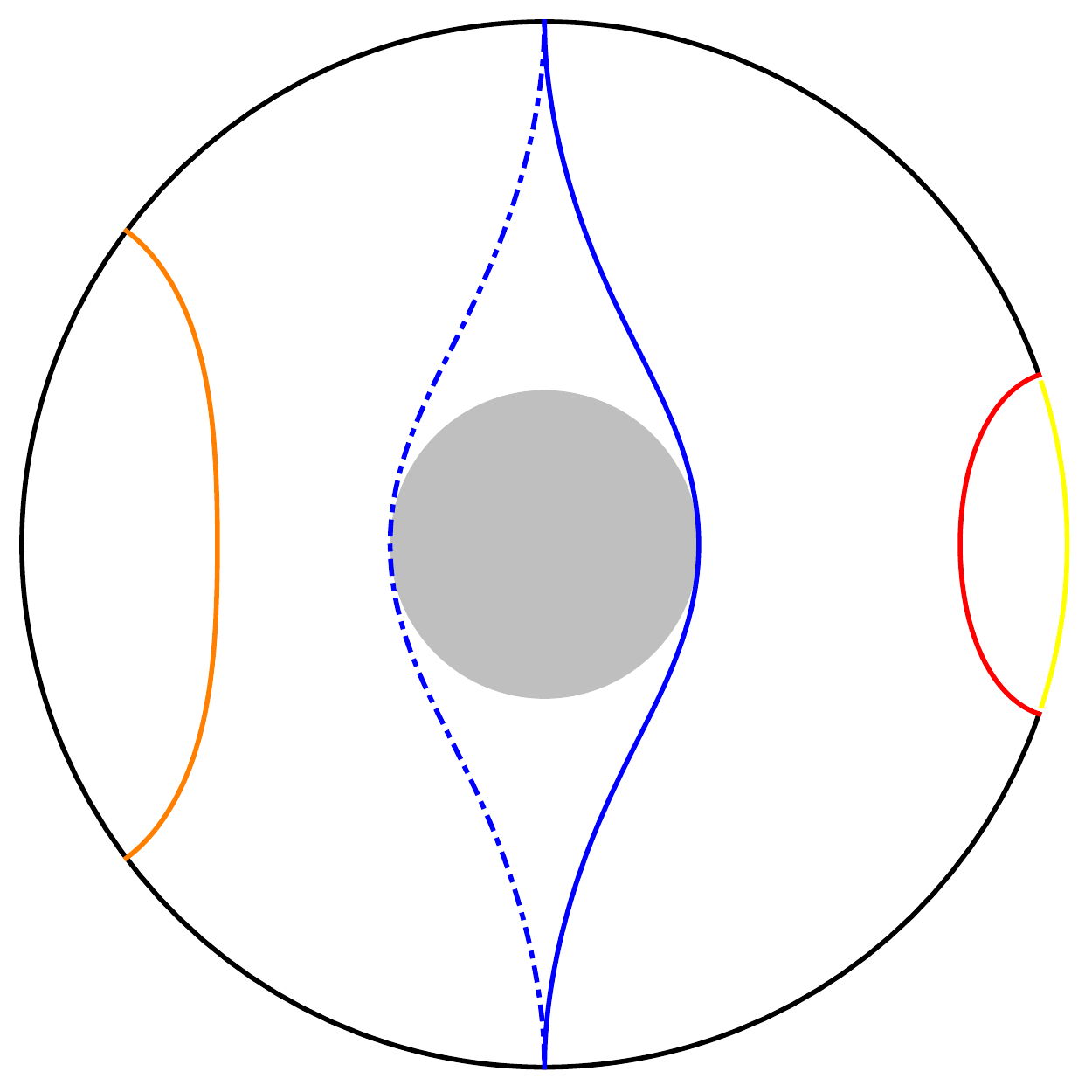}
		\caption{}
		\vskip 2mm
		\label{fig: RTsConical}
	\end{subfigure}
	\begin{subfigure}[b]{0.45\textwidth}
		\centering
		\includegraphics[width=.9\textwidth]{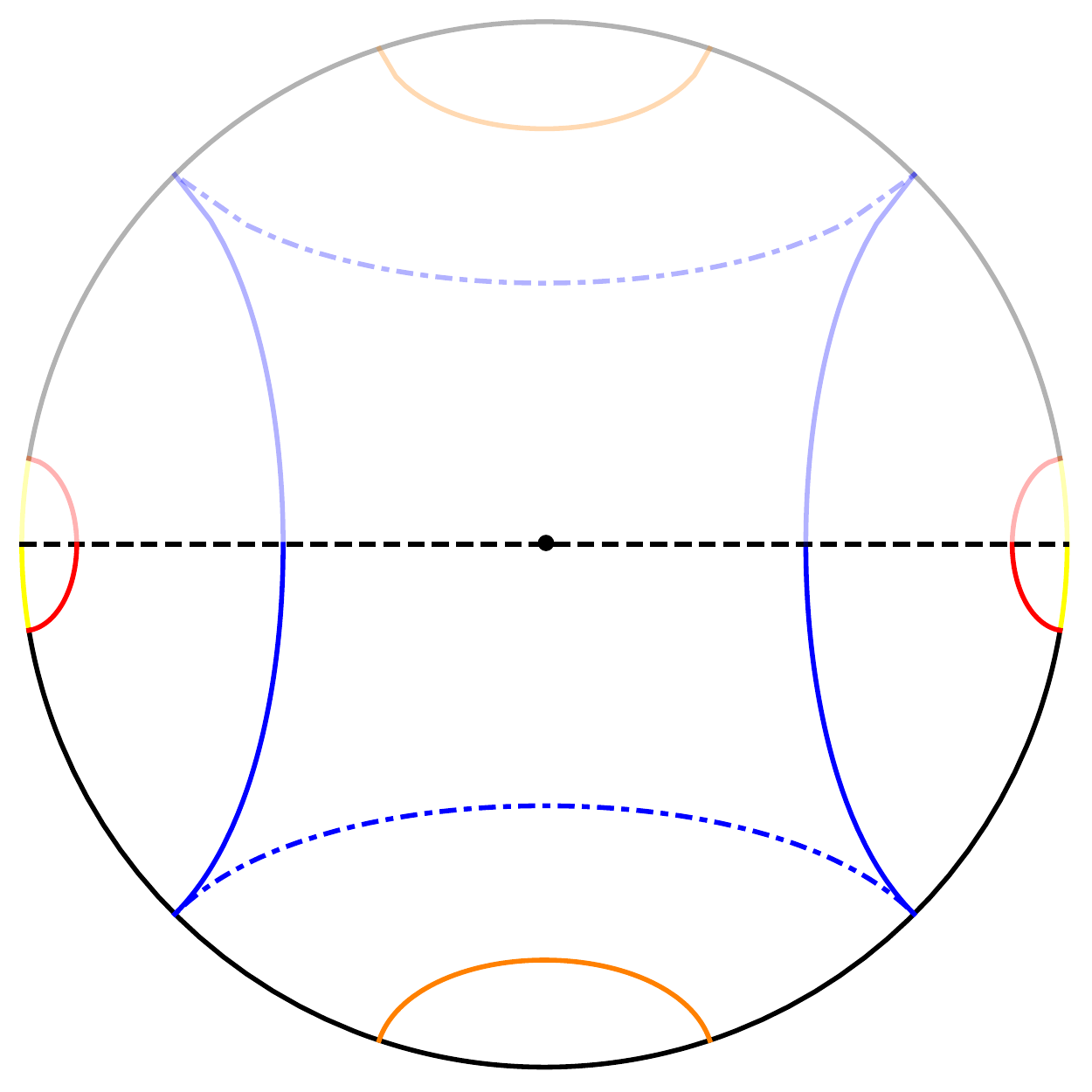}
		\vskip 2mm
		\caption{}
		\label{fig: RTsCovering}
	\end{subfigure}
    \caption{A constant time slice of conical AdS$_3$ with $\alpha=1/2$ is shown. The boundary is pulled to a finite value by using the radial coordinate, $\rho=\arctan r$. The entanglement shadow is shown in gray. A few representative HRT surfaces are shown in orange, blue and red on the fundamental domain in (a). The same geodesics are shown on the covering space (AdS$_3$) in (b) together with their images. The fundamental domain is obtained by identifying points under a rotation by $\pi$.}
    \label{FigHRTSurfCon}
\end{figure}

This characterisation of the spacelike geodesics in AdS$_3$ allows us to construct the HRT surface anchored to a boundary interval. For the conical deficit spacetime, we can similarly construct all of the candidate HRT surfaces by looking at all the surfaces anchored to images of the boundary points at the edges of the boundary region. The true HRT surface is the one of minimal length. This condition causes conical deficit spacetimes to develop an entanglement shadow, a bulk region that no HRT surface can reach, around the conical singularity at $r=0$. The minimal radius probed by the HRT surfaces can be found to be \cite{entwinement14}
\begin{equation}
r_{\text{min}} = \alpha \cot \alpha \frac{ \pi}{2}.
\end{equation}
A few HRT surfaces together with the entanglement shadow are drawn in figure \ref{FigHRTSurfCon} for $\alpha = 1/2$. Any simply connected boundary subregion has the same HRT surface as its complement. The red geodesic in figure \ref{FigHRTSurfCon} can be therefore be thought of as the HRT surface to the black or the yellow subregion of the boundary. Note that the entanglement shadow is included in the homology surface of the black boundary region, in contrast to $H_R$ of the yellow boundary region. 

The presence of this region unprobed by HRT surfaces makes the determination of the localisable region more subtle than in AdS$_3$. \cite{SW} suggested that those entanglement shadows nevertheless belong to the localisable region because, by their theorem III.1, the geometric objects that matter in the determination of the localisable region are entanglement wedges. They argued, based on their figure 3, that the entire spacetime was localisable. This argument is recapped in our figure \ref{fig: assumption wedges}. We will now construct the entanglement wedges in the conical deficit spacetime in order to verify whether the behaviour depicted in this figure in generic.

\subsection{Entanglement wedges}

Given an HRT surface, the entanglement wedge can be obtained by lightsheet construction outlined in \cite{HRT07}. The idea is that from each point on the HRT surface two light rays are shot orthogonally to the HRT surface in the direction of the $H_R$ hypersurface (one future- and one past-directed). The collection of these light rays together with the boundary causal diamond of the HRT surface forms the boundary of the entanglement wedge. In this construction, one should take into account that the lightsheet must be terminated whenever light rays intersect. Such intersections are called caustics. In this section, we will derive the location of the lightsheets and caustic analytically. In Appendix \ref{sec: Numerical approach to the lightsheet construction}, a numerical approach to this construction is summarised for the case of a conical deficit, which is also applicable to other spacetimes.

The light rays generating the lightsheet are null geodesics and so they can be described by a 2-plane in $\mathbb{R}^{2,2}$ spanned by a timelike unit vector $Y(\xi)$ where the null ray leaves the HRT surface and a null tangent vector $N$. This light ray must be orthogonal to the HRT surface so that $N \cdot X_1 =0$. This means that $N$ must live on the 2-plane normal to the HRT surface so that each of the two lightsheets are generated by one of the two null vectors spanning the normal plane, $N_1$ and $N_2$,\footnote{Note that the same two null vectors $N_1$ and $N_2$ generate the normal space along the entire HRT surface. In AdS, the normal space rotates as we move along the HRT surface. However the HRT surfaces lift to planes in $\mathbb{R}^{2,2}$ where parallel transport is trivial. The rotation of the normal space in AdS therefore comes from pulling back these fixed vectors through the map \eqref{eqn:cone_param}. This was discussed in \cite{higher_dim_diff_ent18}. } %
\begin{align}
L_i(\lambda,\xi) = Y(\xi) + \lambda N_i\,, \qquad \mathrm{for}~\lambda>0~\mathrm{and}~i=1,2\,.
\end{align}

For the conical deficit, explicit expressions for this lightsheet can be obtained. We will focus on the entanglement wedges of boundary intervals covering more than half of the boundary, so that the lightsheets will initially point towards decreasing $r$. The construction of the lightsheets of the complementary region (of size smaller than half the boundary) is straightforward and similar, although these regions will not have caustics in the bulk for conical deficits.

We shall parametrise the HRT surface by the location of the point whose future light ray along the lightsheet will hit the conical singularity at $r=0$. Without loss of generality, we can choose coordinates such that this point lies at $t=0$ and $\theta=0$. Therefore,
\begin{align}
X_0 = \left( \frac{\sqrt{r_0^2+\alpha^2}}{\alpha}, 0,0, \frac{r_0}\alpha  \right) \,,
\end{align}
as before. We will now choose a basis for embedding space by pushing forward the tangent space of this point. By pushing forward the unit vectors in each coordinate direction, we can construct the following embedding space vectors
\begin{align}
R &= \sqrt{r_0^2 + \alpha^2} \partial_r X^A_{cone} (r_0,0,0)
= \alpha^{-1} \left( r_0, 0 ,0, \sqrt{r_0^2 + \alpha^2} \right)\,,\\
T &= \frac{1}{\sqrt{r_0^2 + \alpha^2}} \partial_t X^A_{cone}(r_0,0,0)
= \left( 0,1,0,0 \right) \,, \\
\Theta &= \frac1{r_0} \partial_\theta X^A_{cone}(r_0,0,0)
= \left( 0,0,1,0 \right)\,.
\end{align}
Taken together, $(X_0,\,T,\,\,R,\,\Theta)$ form an orthonormal basis for $\mathbb{R}^{2,2}$. Using this basis, we can provide an intuition for the parametrisation of the HRT surfaces that we used in \eqref{eqn:hrt_cone}. The future-directed light ray leaving $X_0$, which will hit the conical singularity, is characterised by the fact that it will leave the HRT surface in a null direction with no angular component. Since the metric has no cross terms or $\theta$ dependence, a geodesic that starts with no velocity in the $\theta$ direction will never acquire one. This inward pointing future-directed null vector, orthogonal to $X_0$, with no angular component can be constructed as 
\begin{align}
N_1 = T - R \,.
\end{align}
We want to construct an HRT surface such that the null ray in this direction will be the generator of the lightsheet leaving from $X_0$. Therefore, this null vector must be one of the $N_i$ characterising the space orthogonal to the $X_0$--$X_1$ plane. We must therefore choose $X_1$ so that it is orthogonal to this vector. A general spacelike unit vector orthogonal to both $X_0$ and $N_1$ can be parametrised by a single parameter $\eta \in \mathbb{R}$,
\begin{align}
X_1 = \Theta + \eta \left( T-R \right)\,.
\end{align}
For $\eta=0$, this tangent vector has no component in the time direction, so the resulting HRT surface will stay at a fixed time. One can therefore think of $\eta$ as parametrising the tilt of the boundary interval away from the constant time slice.

The final null vector completing our orthonormal frame is determined by the requirement that it be (i) orthogonal to $X_0$ and $X_1$, (ii) null and (iii) satisfy $N_1 \cdot N_2 = -2$,
\begin{align}
N_2 =(\eta^2-1) R -(\eta^2+1)T - 2\eta \Theta  \,.
\end{align}

\begin{figure}
	\centering
	\begin{subfigure}[b]{0.45\textwidth}
		\centering
		\includegraphics[width=.75\textwidth]{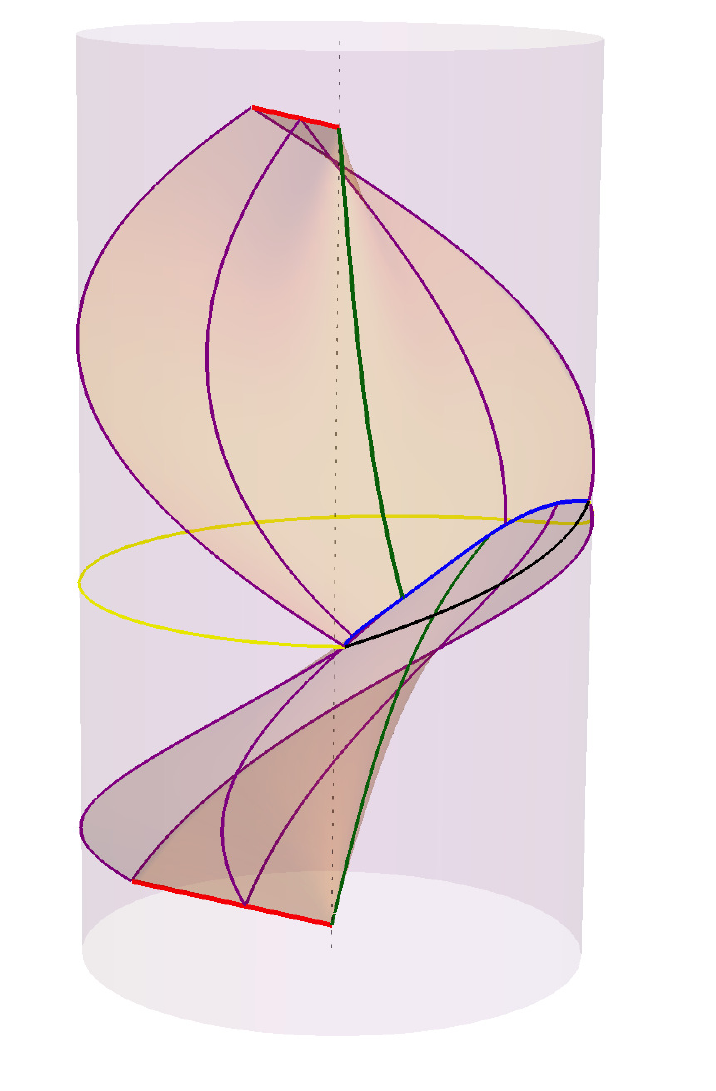}
		\vskip 2mm
		\caption{$\alpha=\frac{1}{2}, r_0=2,\eta=0.5$}
		\label{fig: LightsheetN2a}
	\end{subfigure}
	\begin{subfigure}[b]{0.45\textwidth}
		\centering
		\includegraphics[width=.75\textwidth]{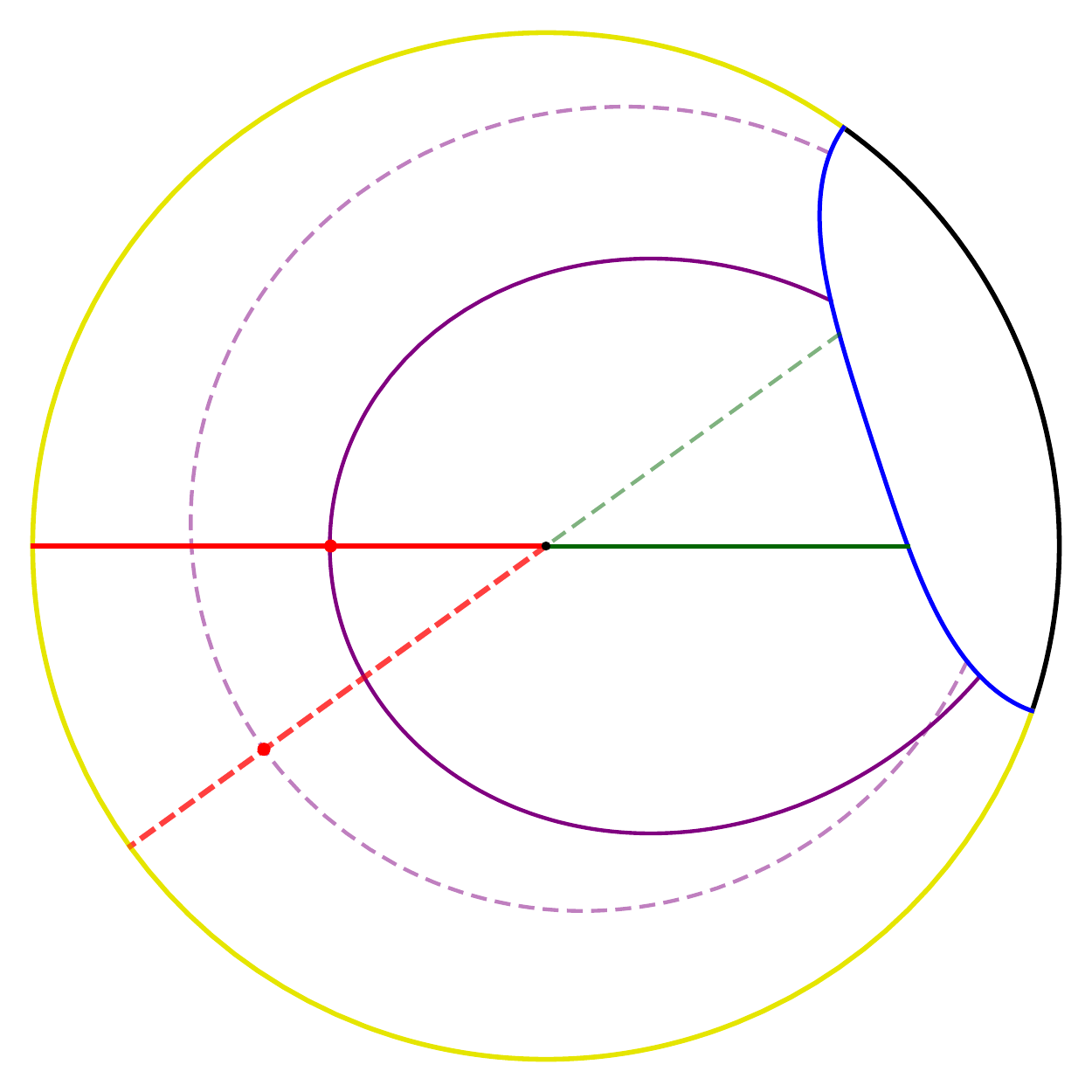}
		\caption{$\alpha=\frac{1}{2}, r_0=2,\eta=0.5$}
		\vskip 2mm
		\label{fig: LightsheetN2b}
	\end{subfigure}
	\begin{subfigure}[b]{0.45\textwidth}
		\centering
		\includegraphics[width=.75\textwidth]{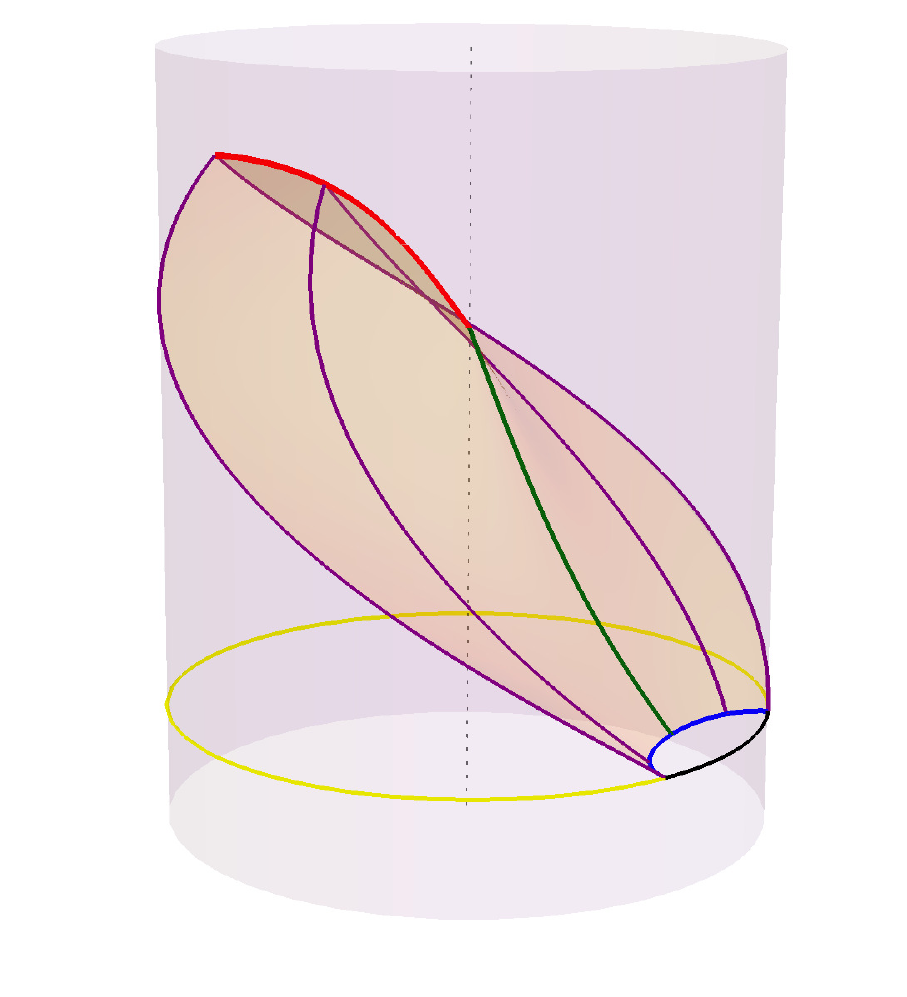}
		\vskip 2mm
		\caption{$\alpha=\frac{2}{3}, r_0=2.5,\eta=0$}
		\label{fig: LightsheetN1_5}
	\end{subfigure}
		\begin{subfigure}[b]{0.45\textwidth}
		\centering
		\includegraphics[width=.75\textwidth]{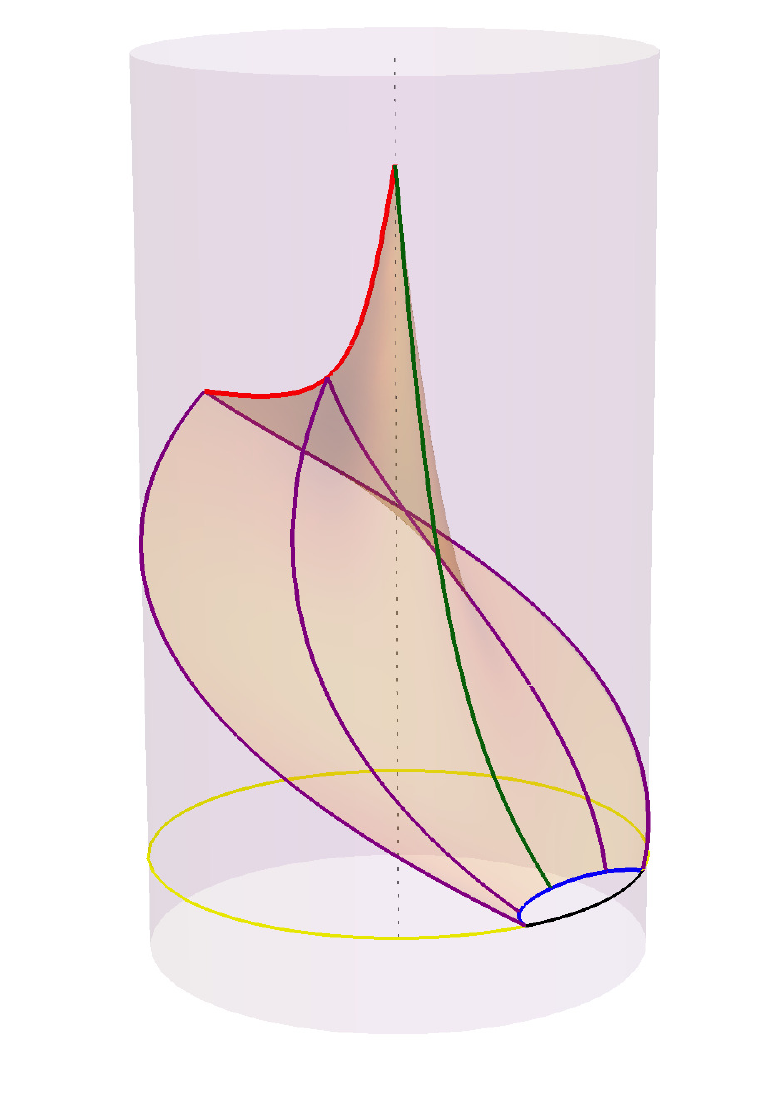}
		\vskip 2mm
		\caption{$\alpha=\frac{1}{3}, r_0=2.5,\eta=0$}
		\label{fig: LightsheetN3}
	\end{subfigure}
\caption{Plots of the lightsheet bounding the entanglement wedges corresponding to intervals, shown in yellow, that cover more than half of the boundary. The HRT surface is displayed in blue. A few light rays generating the lightsheet are drawn in purple. The light ray that hits the conical singularity at $r=0$ is drawn in green and the caustic where the lightsheet terminates is in red.
 In (a) and (b) the front and top view respectively of the full lightsheet which bounds the entanglement wedge of a non-equal time slice HRT surface is displayed, with the dashed lines in (b) referring to the past lightsheet and the solid lines to the future lightsheet. For $\alpha=\frac12$, the caustic is at constant $t$. In (c) and (d) only the future lightsheet is displayed, so as to reduce clutter.}
 \label{fig: Lightsheets}
\end{figure}

The future lightsheet is therefore located at
\begin{align}
L_1 (\xi,\lambda) %
&=\sec \xi \, X_0 + \tan \xi \, \Theta + (\lambda + \eta \tan\xi ) \, T- (\lambda + \eta \tan\xi) \, R    \,,\nonumber\\
&= \Bigg(\frac{\sqrt{r_0^2 + \alpha^2} \sec\xi - r_0 (\lambda + \eta \tan\xi) }{\alpha} 
,\, \lambda + \eta \tan\xi ,\, \label{eqn:lightsheet_cone}\\ 
& \qquad\qquad \tan \xi ,\,  
 \frac{r_0 \sec\xi - \sqrt{r_0^2 + \alpha^2} (\lambda + \eta \tan\xi)}{\alpha} \Bigg)\, ,
\nonumber
\end{align}
and is shown in figure \ref{fig: Lightsheets} for a few cases. 

This lightsheet must be terminated whenever two generators cross, so that $L_i(\lambda_1,\xi_1) = L_i(\lambda_2,\xi_2)$. In the case of AdS$_3$, that is $\alpha=1$, the planes in $\mathbb{R}^{2,2}$ containing these generators intersect only along the null ray $N_i$, which corresponds to the boundary point at the tip of the boundary causal diamond associated to the region on which the HRT surface is anchored. This is the fact that in AdS, the lightsheets are free of caustics in the bulk and terminate at the tip of the boundary causal diamond.

The new ingredient in the conical deficit spacetime is the identification. When the interval covers more than half of the boundary, this leads to new solutions to $L_i(\lambda_1,\xi_1) = L_i(\lambda_2,\xi_2)$ as two vectors can be related by the identification. Recall that this identification corresponds to a rotation by $2\pi\alpha$ of $\phi$ in the hyperpolar coordinates \eqref{eqn:hyperpolar}, which parametrises the angle in the positive signature coordinates of $\mathbb{R}^{2,2}$.

In general, identifying a caustic requires tuning three of the four parameters $(\xi_1,\lambda_1,\xi_2,\lambda_2)$. However, in this case we can exploit symmetries in our set-up to simplify our task. From the explicit expression for $L_1^A$, we see that $L_1^3$ is odd in $\xi$. Since we can apply our identification symmetrically under a $X^3 \rightarrow -X^3$ reflection by identifying the plane $P_{plane}(\pi \alpha)$ with $P_{plane}(-\pi \alpha)$, we should look for caustics where the lightsheets reach these planes. We find a caustic where $L_1(\lambda_1,\xi) \cdot P_{plane}(\pi \alpha) =0$ and $L_1(\lambda_2,-\xi) \cdot P_{plane}(- \pi \alpha) =0$.
This occurs at
\begin{align}
\lambda_1 = - \frac{Y(\xi)\cdot P_{plane}(\pi \alpha) }{N_1\cdot P_{plane}(\pi \alpha) } \,, \qquad
\lambda_2 = - \frac{Y(-\xi)\cdot P_{plane}(-\pi \alpha) }{N_1\cdot P_{plane}(-\pi \alpha) } \,.
\label{eqn:self_intersection_soln}
\end{align}
The caustic is located at 
\begin{align}
L_1(\lambda_1,\xi) &= \left(
\frac{\alpha \sec\xi - r_0 \cot\pi \alpha \tan\xi}{\sqrt{r_0^2 + \alpha^2 }},
\frac{r_0 \sec\xi + \alpha \cot\pi\alpha \tan\xi}{\sqrt{r_0^2 + \alpha^2}},
\tan\xi, -\cot \pi \alpha \tan\xi
\right) \,, \nonumber\\
L_1(\lambda_2,-\xi) &= \left(
\frac{\alpha \sec\xi - r_0 \cot\pi \alpha \tan\xi}{\sqrt{r_0^2 + \alpha^2 }},
\frac{r_0 \sec\xi + \alpha \cot\pi\alpha \tan\xi}{\sqrt{r_0^2 + \alpha^2}},
-\tan\xi, -\cot \pi \alpha \tan\xi
\right) \,.
\end{align}
In terms of the coordinates  covering the conical deficit defined in \eqref{eqn:cone_param}, $L_1(\lambda_1,\xi)$ and $L_1(\lambda_2,-\xi)$ are located at $\theta = \pm \pi$ respectively along the same curve,\footnote{The branch of $\arctan$ which has range $[0,\pi]$ must be used. This branch ensures that $t(r)$ is continuous as $\alpha$ is varied near $\alpha=\frac12$.}
\begin{align}
t(r) 
= \frac1\alpha \left(  \arctan \frac{\sqrt{\alpha^2 + r^2 \sin^2 \pi \alpha}}{r \cos \pi \alpha} 
-\arctan \frac{\alpha}{r0} 
\right) \,,
\label{eq: caustic conical}
\end{align}
confirming that this is a caustic.

There is a simple expression for $\partial_r t(r)$, which makes manifest its definite sign:
\begin{align}
 \partial_r t(r) = -
\frac{\alpha \cos \pi\alpha}{\left(r^2+\alpha ^2\right) \sqrt{\alpha ^2+r^2
   \sin ^2 \pi\alpha}} \,.
   \label{eqn:dtdr_cone_caustic}
\end{align}

\begin{figure}
	\centering
	\begin{subfigure}[b]{0.7\textwidth}
		\centering
		\includegraphics[width=.95\textwidth]{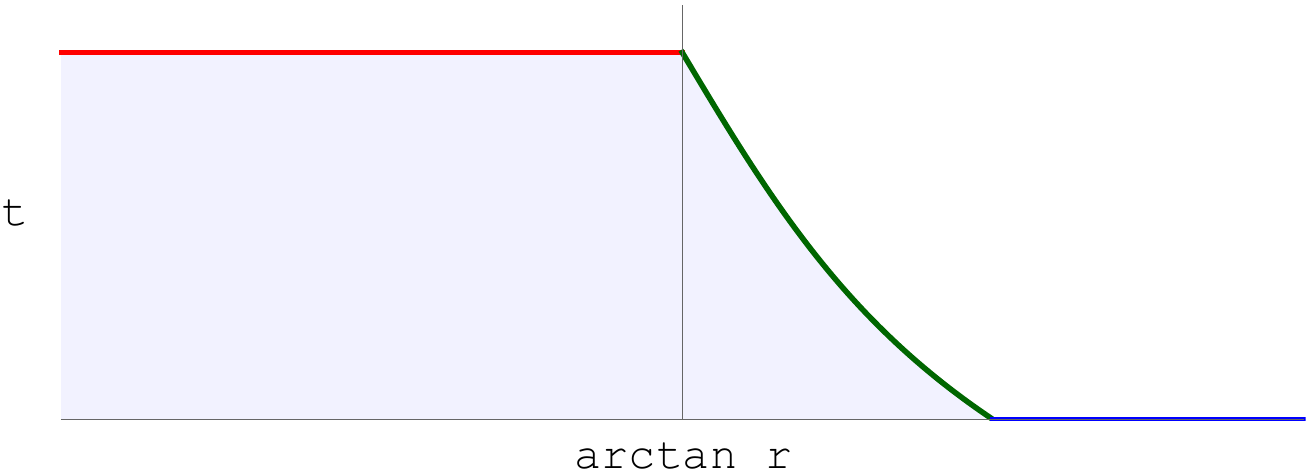}
		\vskip 2mm
		\caption{$\alpha = \frac{1}{2}$}
		\label{fig: caustic2}
	\end{subfigure}
		\begin{subfigure}[b]{0.7\textwidth}
		\centering
		\includegraphics[width=.95\textwidth]{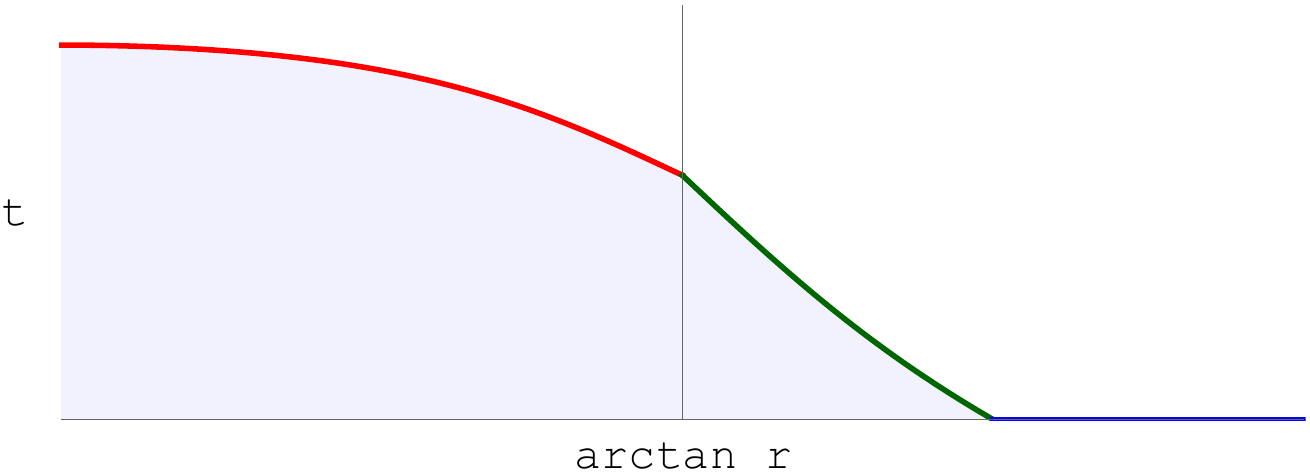}
		\vskip 2mm
		\caption{$\alpha = \frac{2}{3}$}
		\label{fig: caustic1_5}
	\end{subfigure}
	\begin{subfigure}[b]{0.7\textwidth}
		\centering
		\includegraphics[width=.95\textwidth]{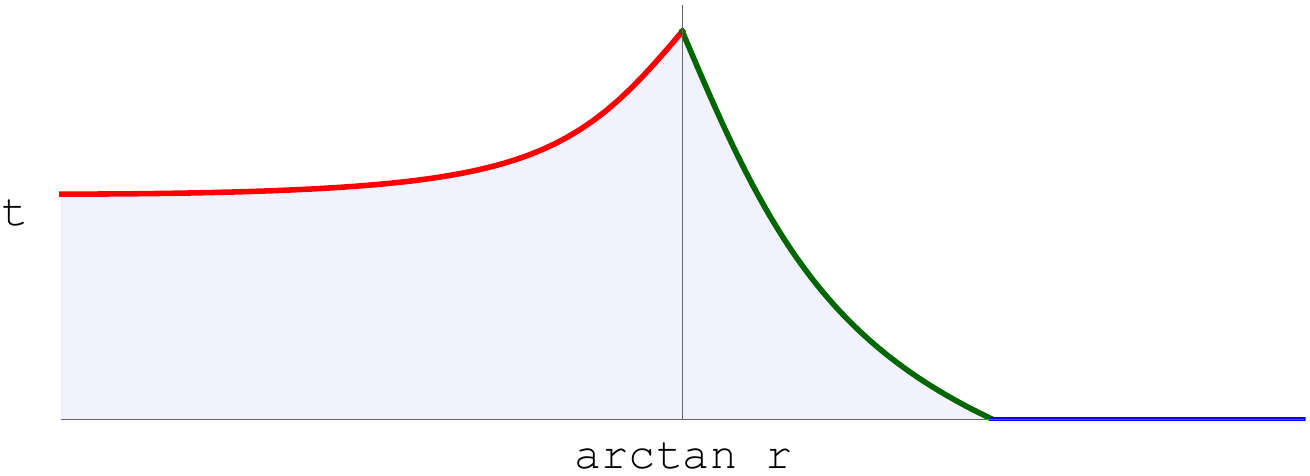}
		\caption{$\alpha = \frac{1}{3}$}
		\vskip 2mm
		\label{fig: caustic3}
	\end{subfigure}
	\caption{A side view of the entanglement wedge for regions covering more than half of the boundary in the conical deficit for various values of $\alpha$. The HRT surface is depicted in blue, the ingoing light ray in green, the caustic in red and the interior of the entanglement wedge is shaded in light blue.  }
    \label{fig: caustics}
\end{figure}

This result has three interesting features. The first is that it does not depend on $\eta$ used to parametrise the tilt of the HRT surface. The caustic leaves the conical singularity at $r=0$ and moves towards the boundary at constant $\theta$. It hits the boundary at the future tip of the boundary causal diamond, at $\theta = \pm \pi$ and  
\begin{align}
t(r=\infty)
= 
\pi  - \frac1\alpha \arctan \frac{\alpha}{r_0} \,.
\label{eq: time boundary tip}
\end{align}
That the caustic does not depend on $\eta$ reflects the fact that its shape does not depend on whether the past tip of the boundary causal diamond is at the same angular position as the future tip. 
In effect, we have chosen our coordinates so that the future caustic and the future tip of the causal diamond all lie at $\theta=\pm \pi$. In these coordinates, the choice of $r_0$ and $\eta$ determine where the past tip will lie. By the time reflection symmetry of the metric, it must be that the caustic on the past lightsheet also lies at the angle of the past tip. Hence for tilted HRT surfaces, the past caustic will not lie at $\theta=\pm \pi$ anymore, as illustrated in figure \ref{fig: LightsheetN2a} and \ref{fig: LightsheetN2b}.

The second feature is that its shape does not depend on $r_0$. The only effect of $r_0$ is to shift the caustic in $t$, as the value of $r_0$ determines the position of the future tip where the caustic meets the boundary.

The last feature is that $t(r)$ is monotonic, since \eqref{eqn:dtdr_cone_caustic} does not change sign as a function of $r$. $t(r)$ is decreasing for $0<\alpha<\frac12$ and increasing for $\frac12 < \alpha <1$, as shown in figure \ref{fig: caustics}. For $\alpha = \frac12$, the caustic is flat since $\partial_r t(r) =0$. Moreover, the difference between the time at which the radial light ray reaches the singularity and the time of the future tip of the boundary causal diamond can be seen to be
\begin{align*}
t(\infty)-t(0)= \pi - \frac{\pi}{2 \alpha} \,.
\label{eq: time difference tip singularity}
\end{align*}

\begin{figure}
	\begin{subfigure}[b]{0.45\textwidth}
	\centering
		\includegraphics[width=.9\textwidth]{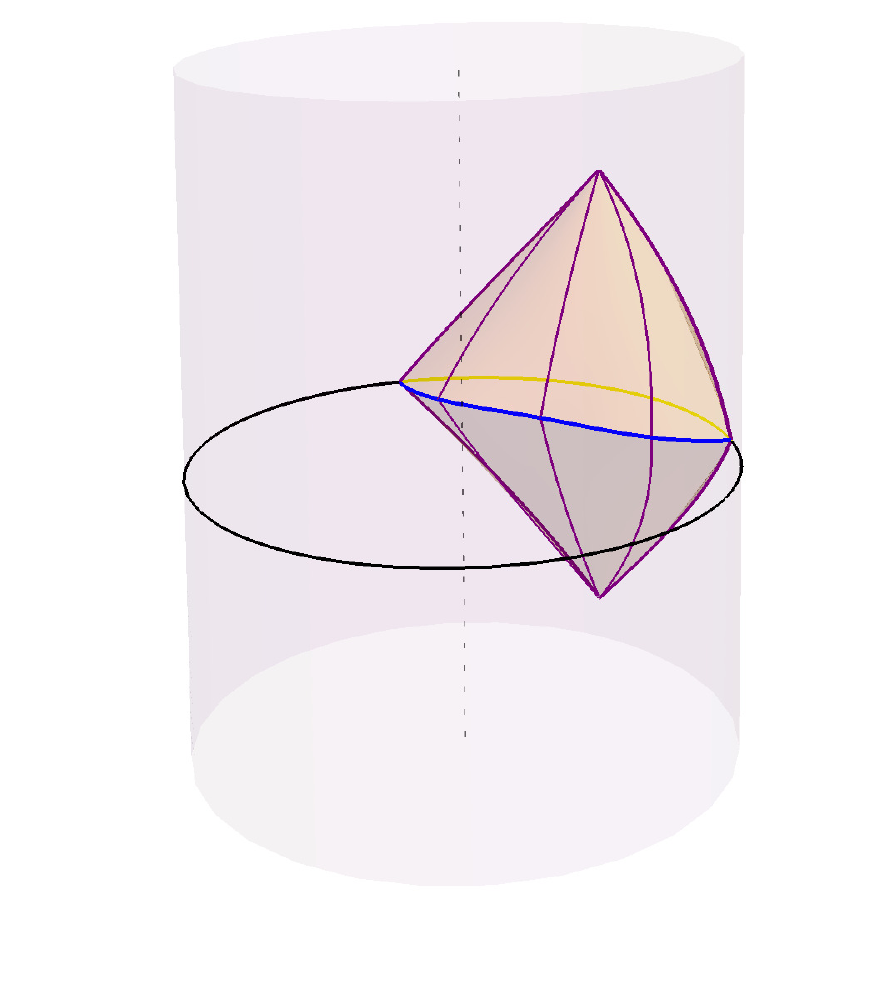}
		\vskip 2mm
	\end{subfigure}
	\begin{subfigure}[b]{0.45\textwidth}
		\centering
		\includegraphics[width=.85\textwidth]{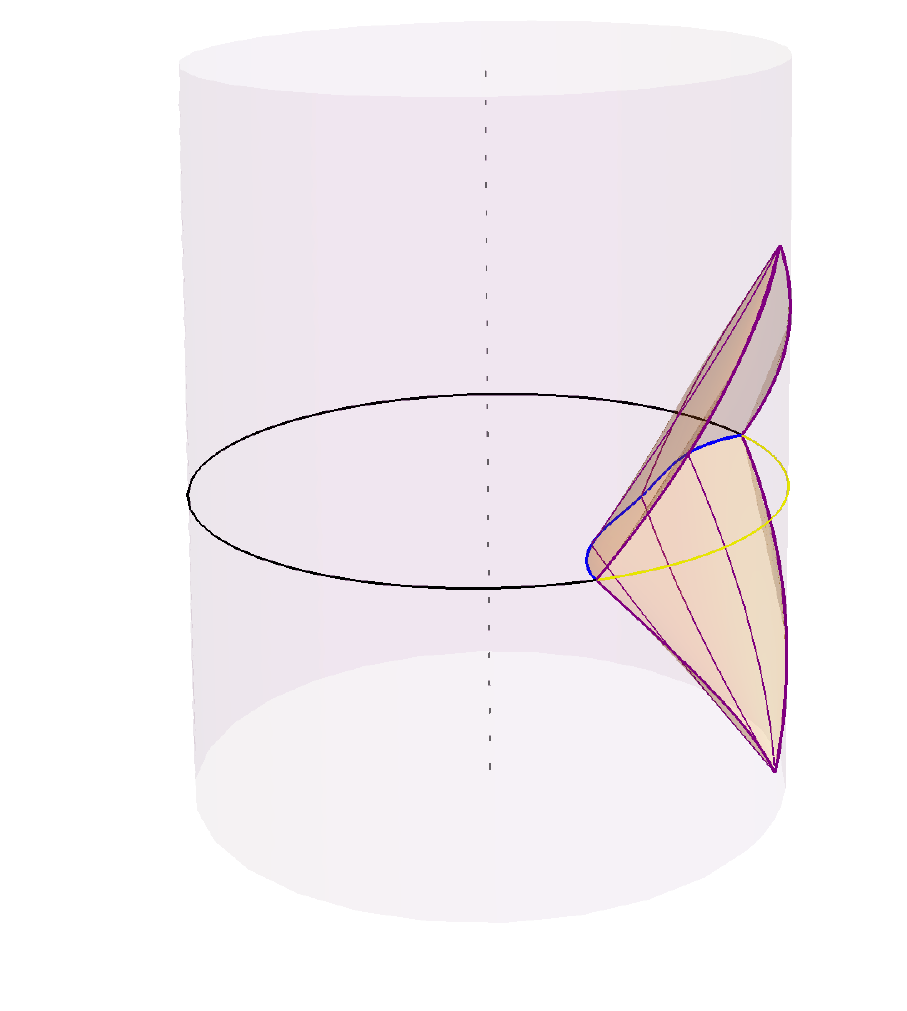}
		\vskip 2mm
	\end{subfigure}
\caption{Plot of the lightsheet bounding the entanglement wedge corresponding to an interval, shown in yellow, that covers less than half of the boundary with $r_0 = 1.1$, $\eta = 0$ and $\alpha = 1/3$. A few light rays generating the lightsheet are drawn in purple.}
\label{fig: entanglement wedges interval less than half}
\end{figure}

Lightsheets of intervals containing less than half of the boundary can be constructed in a similar way and are shown for completeness in figure \ref{fig: entanglement wedges interval less than half}. 

\subsection{Localisable region}

Let us now turn to determining the localisable region in the conical deficit spacetime. The argument given in \cite{SW} for localisability in the conical deficit, reviewed in our figure \ref{fig: assumption wedges}, assumed that $t(r)$ describing the caustics is monotonically increasing, but we have seen that this is not the case for $0< \alpha \leq 1/2$. In particular, this assumption does not hold for any of the conical deficit spacetimes obtained by a $\mathbb{Z}_n$ identification, which have $\alpha = \frac{1}{n} $ \cite{coneZn}. These are the spacetimes where entwinement was proposed as a quantity that could probe inside the entanglement shadow \cite{entwinement14,entwinement16,entwinement18}.

We have found that $t(r)$ is indeed monotonically increasing for conical deficits $\frac12<\alpha<1$, so their argument goes through and we conclude that the entire spacetime is localisable. However, for $0 < \alpha \leq \frac12$ this is not the case. In those cases, we can show that there is a non-localisable region coinciding with the entanglement shadow as follows. Points inside the entanglement shadow are only inside the entanglement wedge of boundary intervals that cover more than half of the boundary. These entanglement wedges are bounded by the radial light ray heading from the HRT surface directly to the conical singularity along a direction $\theta=\theta_0$ and by the caustic at $\theta=\theta_0 \pm \pi$. Since $t(r)$ is monotonically decreasing for both the caustic and the ingoing light ray, these entanglement wedges will always include a whole interval $[0,r]$ at fixed time $t(r)$ and $\theta$. Therefore any entanglement wedge that includes a point $(r_*, t_*, \theta_*)$ along this ingoing light ray or the caustic will also include all the points $(r,t_*,\theta_*)$ with $r<r_*$. By theorem III.1 of \cite{SW}, this implies that the point $(r_*, t_*, \theta_*)$ cannot be localisable.

Although points inside the entanglement shadow are not in the localisable region, the total radial extent of an operator near the conical singularity can be determined from where it can be reconstructed on the boundary.\footnote{We would like to thank Sean Weinberg for emphasising this fact in correspondence on this topic.} The obstruction is that an operator supported on a ring at fixed radius can be reconstructed in exactly the same regions as an operator that is supported on the disk inside this ring.

\subsection{Disconnected boundary regions}

To complete the argument that the entanglement shadow is not in the localisable region for $0 < \alpha \leq \frac12$, we should also analyse regions with multiple disconnected components. Start with the 2 interval case, where the region on the boundary is $R = I_1 \cup I_2$. 

The boundary of this boundary region, $\partial R$, consists of 4 points. The HRT surface must be anchored at these 4 points, and therefore has two components, each consisting of a spatial geodesic connecting two boundary points. There are two possible ways of connecting the 4 boundary points: either the spatial geodesics connect the endpoints of each interval independently or they connect the intervals to each other. In the first case, the homology surface is the union of two disconnected homology surfaces, each corresponding to a homology surface associated to a single interval of less than half the boundary circle. This situation only has the trivial caustics at the tips of the two boundary diamonds. In the second case, the homology surface connects the two intervals across the bulk and includes the central region around the conical singularity. This situation is the more interesting one with non-trivial caustics.

\begin{figure}
	\centering

	\begin{subfigure}[b]{0.45\textwidth}
	\centering
		\includegraphics[width=1\textwidth]{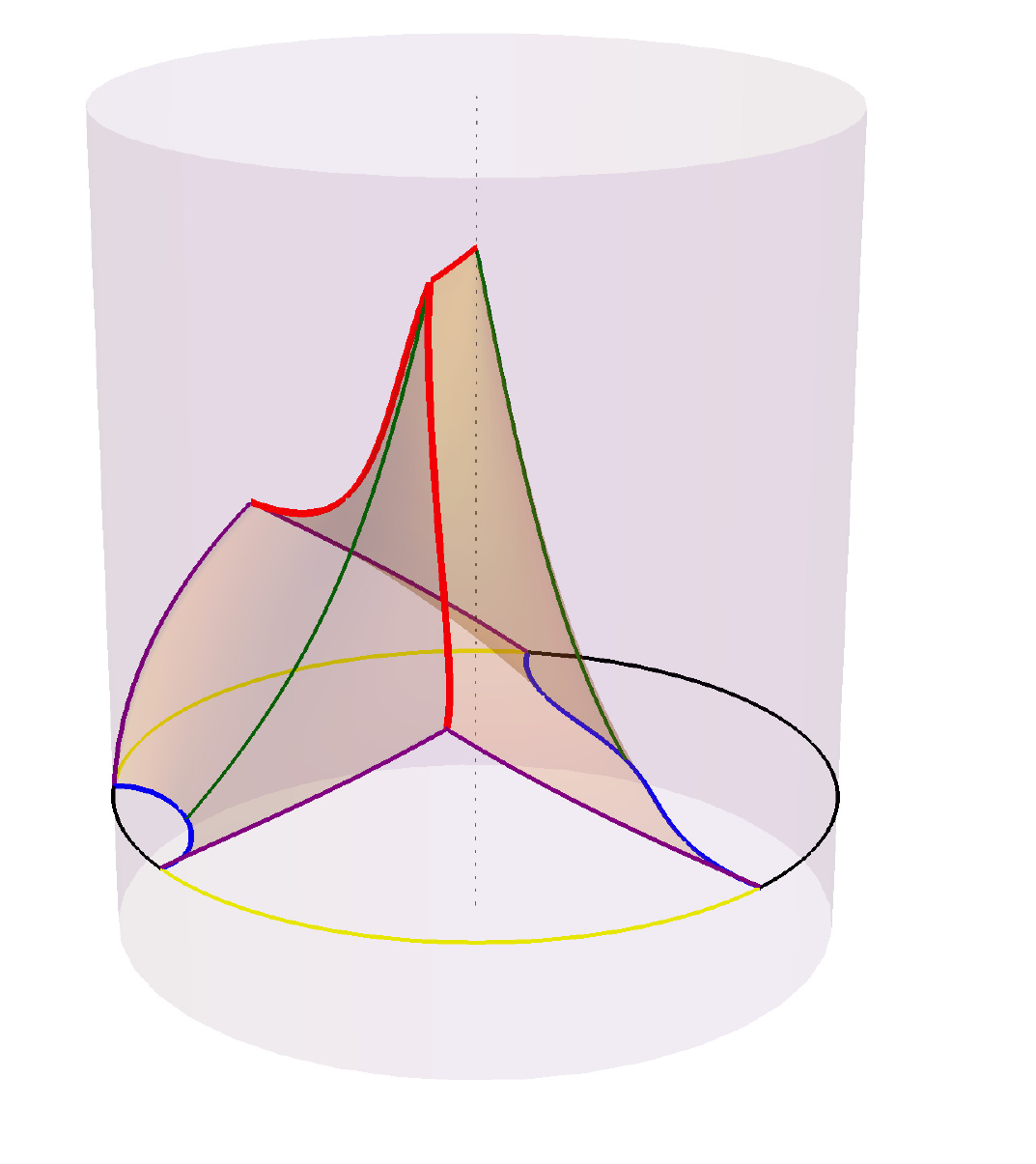}
		\vskip 2mm
		\label{fig: two_intervals_0_85_3_5}
	\end{subfigure}
	\begin{subfigure}[b]{0.45\textwidth}
	\centering
	\includegraphics[width=0.85\textwidth]{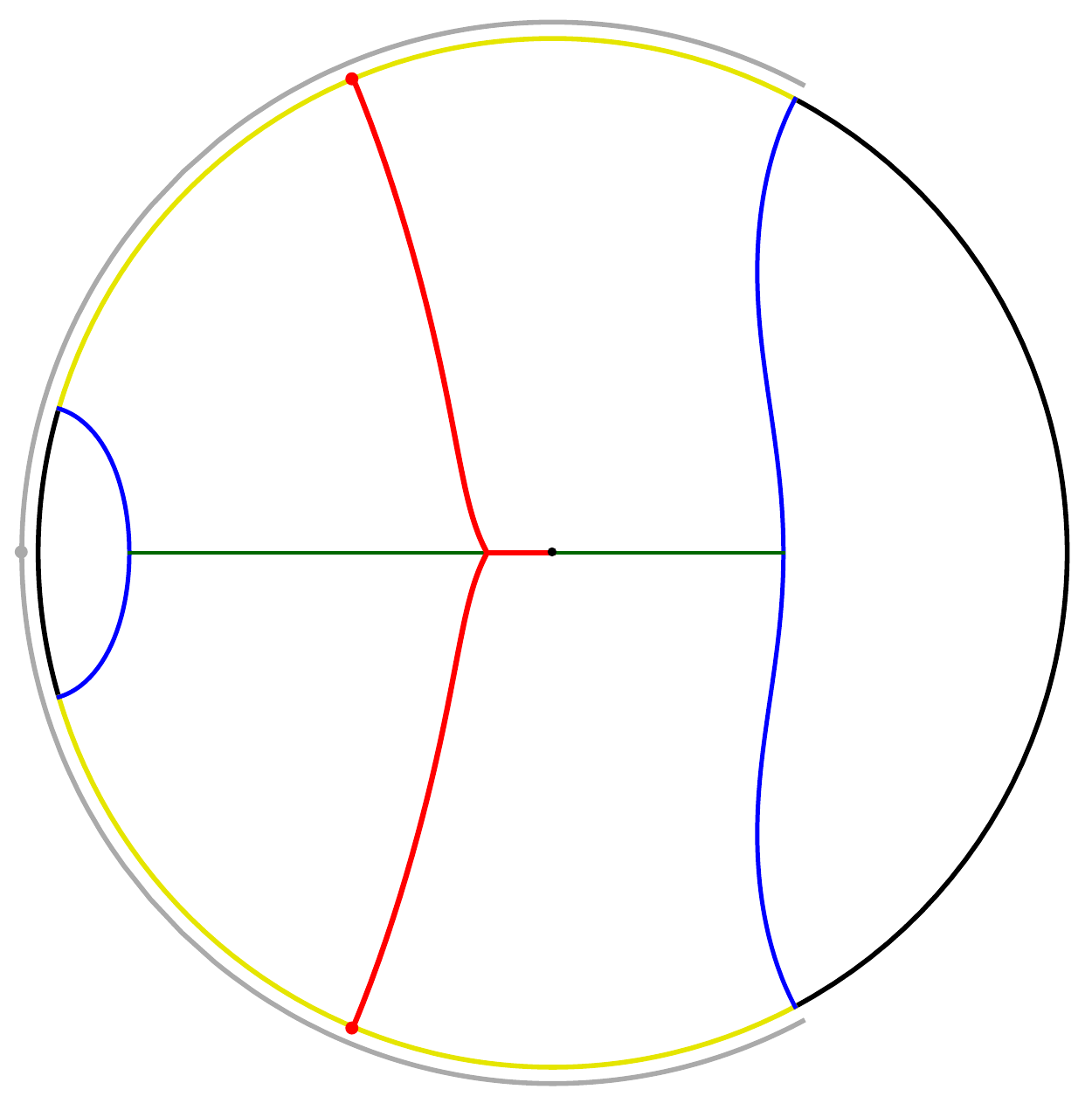}
		\vskip 2mm
		\label{fig: two_intervals_top}
	\end{subfigure}    
	\caption{The entanglement wedge of a disconnected boundary region $R$, shown in yellow, on the constant time slice $t=0$, for  $\alpha = \frac{1}{2.2}$. The HRT surfaces are at $r_0^{(a)}=0.8$, $r_0^{(b)}=3.5$, $\theta_0^{(a)}=0$, $\theta_0^{(b)}=\pi$ and are shown in blue. The caustic, depicted in red, forms a Y-shape that is monotonically decreasing in time as a function of $r$. The green lines represent the two radial light rays starting from the HRT surfaces and meeting the caustic. The purple lines bound the future boundary causal diamonds. \textit{Left.} The complete entanglement wedge seen from the side. \textit{Right.} The caustic seen from the top. The gray segment is a boundary region $R'$ associated to  $\gamma^{(a)}$ alone. The position of the tips of the boundary causal diamonds for the two interval boundary region are illustrated with red dots while the gray dot shows the position of the tip of the single causal diamond of region $R'$.}
	    \label{fig: two_intervals caustic}
\end{figure}

The caustic in this situation is depicted in figure \ref{fig: two_intervals caustic} and can be seen to form a Y-shape. It generically starts at the conical deficit and moves outwards until it splits into two branches, one going to each of the future tips of the two boundary diamonds. A first branch gets formed by the lightsheet emanating from the HRT surface closest to the singularity, where light rays near the radial generator of the lightsheet meet on the other side of the deficit, much as in the single interval case. The other branches come from where the second lightsheet meets this one. The first branch follows exactly the analysis in the previous sections, with the appropriate HRT surface connecting the pair of boundary points that are further apart. The other two branches can be found by looking for the intersection of the lightsheets.

Denote the two HRT surfaces and lightsheets by $(a)$ and $(b)$, where $(a)$ is the one which meets the deficit first and leads to the first branch of the caustic. In the ambient $\mathbb{R}^{2,2}$, the two HRT surfaces are parametrised by 
\begin{align}
Y^{(a)}(\xi^{(a)}) &= \sec \xi^{(a)} X_0^{(a)} + \tan \xi^{(a)} X_1^{(a)} \,, \\ 
Y^{(b)}(\xi^{(b)}) &= \sec \xi^{(b)} X_0^{(b)} + \tan \xi^{(b)} X_1^{(b)} \,,
\end{align}
and the lightsheets are 
\begin{align}
L^{(a)} (\xi^{(a)},\lambda^{(a)}) &= Y^{(a)}(\xi^{(a)}) + \lambda^{(a)} N_1^{(a)}\,, \\ 
L^{(b)} (\xi^{(b)},\lambda^{(b)}) &= Y^{(b)}(\xi^{(b)}) + \lambda^{(b)} N_1^{(b)}\,.
\end{align}
Notice that these lightsheets are simply the intersection of the 3-plane generated by $(X_0,X_1,N_1)$ with the AdS hyperboloid $X^2=-1$. Therefore, the new branch of the caustic will occur along the intersection of these 3-planes. The 3-plane generating the lightsheet is specified by
\begin{align}
N_1 \cdot L =0\,.
\end{align}
Therefore the intersection of the lightsheets occurs when 
\begin{align}
N_1^{(a)} \cdot L^{(b)}(\xi^{(b)},\lambda^{(b)}) =0 \,,
\qquad \mathrm{or} \qquad
N_1^{(b)} \cdot L^{(a)}(\xi^{(a)},\lambda^{(a)}) =0 \,.
\end{align}
These two conditions are the same and must be satisfied at the same points in $\mathbb{R}^{2,2}$, so whichever is more convenient can be used. These conditions are easily solved in terms of the parameter along each generator where this intersection can occur,
\begin{align}
\lambda_*^{(a)} = - \frac{N_1^{(b)} \cdot Y^{(a)}(\xi^{(a)})}{N_1^{(b)} \cdot N_1^{(a)} }\,,
\qquad \mathrm{and} \qquad
\lambda_*^{(b)} = - \frac{N_1^{(a)} \cdot Y^{(b)}(\xi^{(b)})}{N_1^{(a)} \cdot N_1^{(b)} } \,.
\label{eqn:lightsheet_intersection}
\end{align}
The last step is to figure out whether each generator is terminated first by crossing the opposite lightsheet or intersecting with an image of the same lightsheet under the identification required to produce the conical deficit. This amounts to combining the correct branches of the solutions to \eqref{eqn:self_intersection_soln} and \eqref{eqn:lightsheet_intersection}. When doing so, we have implemented the effects of the conical deficit by considering all of the relevant images.

Let us now return to the question of whether the entanglement shadow near the conical deficit is in the localisable region. Since no HRT surfaces pass through the entanglement shadow, the only possibility for localisation is if a region whose entanglement wedge includes the conical singularity has a caustic on the future lightsheet with increasing $t(r)$. This caustic departs the singularity where the first light ray on one of the lightsheets meets it. We denoted by $\gamma^{(a)}$ the HRT surface which emitted this light ray. One can also identify a single interval, $R'$, such that $\gamma^{(a)}$ is its HRT surface and such that its entanglement wedge also includes the conical singularity, as shown in figure \ref{fig: two_intervals caustic}. The lightsheet bounding the entanglement wedge of $R'$ also includes this same light ray that hits the conical singularity. We saw that $t(r)$ parametrising the caustic on the lightsheet of $R'$ was decreasing since, from \eqref{eq: time difference tip singularity}, the time of the future tip of the boundary causal diamond associated to $R'$, $t_{R'}(r=\infty)$, was earlier than $t_{R'}({r=0})$ where the light ray hit the conical singularity,
\begin{align}
t_{R'}(r=0) \geq t_{R'}(r=\infty)\,.
\end{align}
Since $R \subset R'$, the time of the future tips of the causal diamonds associated to $I_1$ and $I_2$, must be less than $t_{R'}(r=\infty)$. 
We therefore expect the branches of the caustic connecting these tips to the branch starting at the conical singularity at $t_{R'}({r=0}) = t_{R}({r=0})$ to be decreasing. In any case, the behaviour right near the conical deficit is controlled by the branch of the caustic that matches that found in the single interval case. Therefore any entanglement wedge constructed in this way includes the same points in the near deficit region and they are not useful in a family of wedges that localises a point through \eqref{eqn:SW_condition_localisability}.

More boundary regions leads to more richness in the possible caustics, but it seems unlikely to us that they will allow us to localise points inside the entanglement shadow. The behaviour of the entanglement wedge near the conical singularity will always be controlled by the first lightsheet to reach it wrapping around the deficit. This leads to the sharp corners we have observed which impede localisability. We can also see that adding more boundary regions will only force the tips of the boundary diamonds, where the caustics must reach the boundary, to earlier times which is not conducive to the type of geometry required to localise new bulk points. 

\subsection{Causal reconstruction in the conical deficit spacetime}
It is interesting to note that causal reconstruction in the conical deficit spacetime also behaves differently for $0<\alpha \leq \frac12$, where the non-localisable region appears, than for $\frac12 <\alpha \leq 1$ where the central region is localisable.

It should first be emphasised that the conical deficit spacetime has no horizons, so that the entire interior can be reconstructed using causal methods when we have access to the entire boundary \cite{BDHM98,HKLL06,dual_density_matrix12,BLR_CW12,HR_CW12}. Since we have an example of a spacetime without a horizon but with a non-localisable region, this demonstrates that being in the localisable region cannot be a necessary condition for whether a local operator can be reconstructed in the boundary theory.
 However, the conical deficit spacetime for $0<\alpha \leq \frac12$ does exhibit a certain type of fragility towards causal reconstruction: omitting even a point from the boundary region means that the causal wedge will no longer include a region around the conical singularity. On the other hand, for $ \frac12 < \alpha \leq 1$, the causal reconstruction of the central region is robust in the sense that the causal wedge corresponding to omitting a single point from the boundary still includes an open region around the conical singularity.

This can be diagnosed by studying a light ray departing from the conical singularity and seeing how long it takes to reach the boundary. The causal diamond corresponding to the entire boundary minus a point terminates at $t = \pi$, where the boundary light rays emitted from the omitted point cross at the other side of the boundary circle. In order for the region near the conical singularity to be reconstructible using causal methods, a light ray departing from it must reach the boundary at $t<\pi$ so that it stays within this causal wedge. A radial outgoing light ray starting at $(r_0,t_0,\theta_0)$ in the conical deficit spacetime follows
\begin{align}
t(r) = t_0 +  \frac{1}{\alpha} \left( \arctan \frac{\alpha}{r_0} - \arctan \frac{\alpha}{r} \right) \,.
\end{align}
Setting $r_0=0$ and $t_0=0$, we see that a radial light ray departing from the conical singularity reaches the boundary at a time $t= \frac{\pi}{2 \alpha}$ confirming the picture discussed above.

\section{BTZ black hole}
\label{sec:two-sided}
In this section we will consider localisability in the BTZ black hole. Localisability in two-sided eternal black holes was considered by \cite{SW} and our analysis will confirm their results. We start by proving a lemma valid in any number of dimensions which provides a sufficent condition for identifying non-localisable regions inside entanglement shadows behind horizons. Turning to the case of the 3-dimensional BTZ black hole, we will find the caustics bounding the entanglement wedges of regions comprising the entirety of one boundary in addition to part of the other. Since these caustics do not impede the innermost light ray from reaching the singularity, the picture form \cite{SW} goes through unchanged. We will then comment on localisability in the one-sided BTZ, where we will conclude that the entanglement shadow is non-localisable.

\subsection{Localisability of entanglement shadows behind horizons}
In the conical deficit spacetime, \cite{SW} proposed a technique, that was reviewed in figure \ref{fig: assumption wedges}, for localising points that cannot be reached by HRT surfaces but that lie on the intersection of lightsheets approaching the point from a future and a past direction.  In this section, we will prove that a region cannot be localised if there are no HRT surfaces in its future light-cone. This provides a connection between regions which are not probed by extremal surfaces, $S$, and localisability in the sense of \cite{SW}: a neighbourhood $U \subset M$ such that $J^+(U) \subset S$ is not localisable.

Many spacetimes are known to have regions that are not probed by extremal surfaces \cite{plateaux13,entwinement14,shadows14}. However, the region near an asymptotically AdS boundary will always be probed by extremal surfaces attached to small boundary regions.\footnote{See for example \cite{EE_from_1st_13} for a discussion of surfaces attached to such small regions.} This means that $S$, the region not probed by extremal surfaces a.k.a. the entanglement shadow, cannot reach the asymptotic boundary. If the future of a neighbourhood is to be contained within the entanglement shadow $S$, and therefore not reach the asymptotic boundary, the spacetime must contain a horizon. Our lemma therefore applies to spacetimes with event horizons, although as we saw in section \ref{sec:cone} in the conical deficit spacetime, event horizons are not necessary for the existence of a non-localisable region.

\begin{lemma}
\label{thm:need-HRT}
Let $U \subset M$ be an open neighbourhood of $M$ such that $J^+(U) \cap \gamma_R=\emptyset$ for all boundary subregions $R$. 
Then $U \subset Loc(M)^c$: this neighbourhood is not localisable.  
\end{lemma}
Proof. We will argue by contradiction. Suppose there exists $p \in U$ that is localisable. Theorem III.1 from \cite{SW} tells us that this is true if and only if there is a family of boundary regions, $\mathcal{R}_0$, such that
\begin{gather}
\bigcap_{R\in \mathcal{R}_0} W_{\mathcal{E}}(R) = \{p\}\,.
\end{gather}

Now consider another point $q\in U\cap J^-(p)$, $q\neq p$. This intersection must be non-empty since $U$ is open. 
Since no HRT surface can intersect the future of $q$, any of the HRT surfaces, $\gamma_R$, anchored to a region $R\in \mathcal{R}_0$ must either enter the past of $q$ or else be entirely spacelike separated from $q$. In either case, it is possible to choose the Cauchy slice of the bulk, $\Sigma_R$, which the HRT surface $\gamma_R$ separates into $H_R$ and $H'_R$, such that $q$ lies to the future of $\Sigma_R$.\footnote{See for example \cite{maximin12} for a discussion of the freedom in choosing this Cauchy slice.}

$p \in W_{\mathcal{E}}(R)$ implies that $p\in D(H_R)$. In fact $p\in D^+(H_R)$, since $p$ is in the future of $q$ and therefore $p$ must also be the future of $H_R \subset \Sigma_R$.
But then, any past-directed causal curve starting at $q$, $\Gamma_q^-$, could be continued to the future along a causal curve connecting $q$ to $p$. Since any inextensible past-directed causal curve through $p$ must cross $H_R$ ($p\in D^+(H_R)$), any such $\Gamma_q^-$ must cross $H_R$ as well. This means that $q\in D^+(H_R)$ and so that $q\in W_{\mathcal{E}}(R)$ for all $R\in \mathcal{R}_0$, in contradiction to the assumption that $p$ is localisable. $\Box$

By simply inverting future and past we can prove another lemma.
\begin{lemma}
\label{thm:need-HRT-past}
Let $U \subset M$ be an open neighbourhood of $M$ such that $J^-(U) \cap \gamma_R=\emptyset$ for all boundary subregions $R$.  Then $U \subset Loc(M)^c$.
\end{lemma}

Thus we see that entanglement shadows provide an obstruction to localisability if they include the entire future or past of a region.  The technique proposed by \cite{SW} and depicted in figure \ref{fig: assumption wedges}, for localising points inside entanglement shadows requires that both the future and the past of the point in question reach outside the entanglement shadow. These lemmas show that this is necessary.

\subsection{Entanglement wedges in the BTZ black hole}
To identify the entanglement wedges and hence the non-localisable region of BTZ, we will
use a similar approach to the previous section on the conical deficit spacetime and describe it as a quotient of AdS$_3$. For the case of non-rotating BTZ, the identification required for taking this quotient can be obtained by an identification of the ambient $\mathbb{R}^{2,2}$, which once restricted to the AdS hyperboloid gives the correct identification. This will allow us to again obtain a closed form expression for the location of the caustic bounding the relevant entanglement wedges.

The identification required to obtain BTZ is most easily described in different (hyperbolic) hyperpolar coordinates on $\mathbb{R}^{2,2}$ of the form
\begin{align}
\left( r_1\sinh\tau, r_2 \cosh\mu , r_1 \cosh\tau, r_2 \sinh\mu \right)\,,
\label{eqn:hyperpolar_BTZ}
\end{align}
where the required identification is $\mu \sim \mu + 2\pi R$. $R$ is the horizon radius in the resulting BTZ measured in units where $L=1$. 

This identifies the plane at $X\cdot{P_{plane}}(\mu_0) =0$ with that at $X\cdot P_{plane}(\mu_0+2\pi R) =0$ where
\begin{align}
P_{plane}(\mu_0) = \left( 0, \sinh\mu_0 , 0, \cosh\mu_0 \right)\,.
\end{align}
Note that the identification required to describe rotating BTZ has a more complicated form and it is not immediately obvious that there is a simple identification of embedding space that restricts correctly to the $X^2=-1$ hyperboloid to reproduce it.

A fundamental domain of this quotient can be covered by coordinates $(u,v,\theta)$,
\begin{align}
X_{BTZ}^A(u,v,\theta) = \left( \frac{v+u}{1+ uv}, \frac{1-uv}{1+uv} \cosh(R \theta),
 \frac{v-u}{1+ uv} , \frac{1-uv}{1+uv} \sinh(R \theta)   \right)\,.
  \label{eqn:BTZcoords}
\end{align}
The metric induced from this embedding is the BTZ metric in Kruskal-like coordinates\footnote{The BTZ black hole was introduced in \cite{BTZ92}. The embedding of BTZ into $
\mathbb{R}^{2,2}$ using these coordinates is reviewed in 
\cite{butterfly13}.
}
\begin{gather}
ds^2 = dX_{BTZ} \cdot dX_{BTZ} = \frac{-4 du dv +R^2 (1-uv)^2 d\theta^2}{(1+uv)^2} \,.
\end{gather}
In these coordinates, the singularity is at $uv=1$ and the right exterior region is $u<0$ and $v>0$.
The boundary is located at $1+uv=0$. A time coordinate can be introduced so that $u=-e^{-R t}$ and $v=e^{R t}$ at the boundary, which is associated to the null rays
\begin{align}
Z_{BTZ}^A (t,\theta) &\propto \frac{1+uv}{2} X^A_{BTZ}(u,v,\theta) \big|_{u=-e^{-R t},\, v=e^{R t}}\\ 
&= \left( \sinh R t , \cosh R \theta,
 \cosh R t ,\sinh R \theta   \right)\,. \label{eqn:BTZ_bdy}
\end{align}

Now we wish to identify the entanglement wedges associated to two types of regions: connected regions contained in the right boundary, as well as the complement of this type of region, which includes the entirety of the left boundary plus a part of the right asymptotic region. Denote the region of interest by $A$. In either case, the corresponding HRT surface is anchored to the right boundary at $\partial A$ and the entanglement wedge is bounded by the radially outward or inward pointing lightsheets respectively from the HRT surface. Regions contained entirely in the right boundary will have HRT surfaces that stay within one fundamental domain of the identification, so they will not develop any new caustics beyond the one at the tip of the boundary diamond. We will therefore focus mostly on the complement type regions. The future-directed lightsheets associated with these regions depart the HRT surface in the $\partial_u$ direction and the past-directed one towards $-\partial_v$. We will focus on the future-directed lightsheet in what follows. As was the case in the last section, the past-directed lightsheet can be understood by exploiting the time reflection symmetry of this metric.

The lightsheet is obtained by following null geodesics orthogonal to each point on the HRT surface to generate a co-dimension 1 surface. Similar to our experience with the conical deficit spacetime, since the metric is rotationally invariant and has no $d\theta$ cross terms, a null geodesic that leaves the surface with no $\partial_\theta$ component to its velocity will fall directly into the singularity in a radial direction. The most important question will then be whether this null generator continues until it hits the singularity or whether nearby generators are bent inwards to cross this ray and form a caustic before this can happen.
We will again label the HRT surfaces by the point which emits this radial light ray. 

This point is described by a vector $X_0(u_0,v_0,\theta_0)$ in the form of \eqref{eqn:BTZcoords}, such that $X_0^2=-1$. We can set $\theta_0=0$ by using the rotational symmetry. The tangent space of the BTZ spacetime can be embedded in embedding space by pushing it forward through the map in \eqref{eqn:BTZcoords}. Similar to the approach taken in the last section, the image of the vectors $\partial_u$, $\partial_v$ and $\partial_\theta$ along with $X_0$, can be normalised to produce an orthonormal frame for the embedding space $(X_0, U,V, \Theta)$,
\begin{align}
X_0&=\Big(\frac{u_0+v_0}{1+u_0 v_0},\frac{1-u_0 v_0}{1+u_0 v_0},\frac{v_0-u_0}{1+u_0
   v_0},0\Big) \,,\\
U&=\Big(\frac{1-v_0^2}{1+u_0 v_0},\frac{-2 v_0}{1+u_0 v_0},\frac{-1-v_0^2}{1+u_0 v_0},0\Big)\,,\\
V&=\Big(\frac{1-u_0^2}{1+u_0 v_0},\frac{-2 u_0}{1+u_0 v_0},\frac{1+u_0^2}{1+u_0 v_0},0\Big)\,,\\
\Theta&=\Big(0,0,0,1\Big)\,. \label{eqn:BTZ_frame}
\end{align}

We now repeat the approach used in the previous section for determining the lightsheet in terms of the ambient $\mathbb{R}^{2,2}$. The first orthogonal null vector defining HRT surface must be chosen to point in the $\partial_u$ direction. This means that $N_1 = U$.

Now we must determine $X_1$ and $N_2$. We will use a similar parametrisation where 
\begin{align}
X_1 = \Theta + \eta U, \qquad N_2 = -2 \eta \Theta - V - \eta^2 U\,,
\end{align}
so that $\eta=0$ describes the surface lying on a constant time slice and $\eta\in \mathbb{R}_0$ describes a boosted or tilted surface.

The resulting HRT surface
\begin{align}
Y(\xi) = \sec \xi X_0 + \tan \xi X_1 \,,
\end{align}
is obtained by imposing $Y^2=-1$ within the $X_0$--$X_1$ plane.

The lightsheets are given by
\begin{align}
L_i(\lambda,\xi) &= Y(\xi)+\lambda N_i \,, \\
L_1 (\lambda,\xi) &= \bigg( 
\frac{\left(u_0+v_0\right)\sec\xi  
+\left(1-v_0^2\right) (\lambda+\eta  \tan\xi)}{1+u_0 v_0}, 
 \frac{ (1-u_0 v_0) \sec\xi -2 \eta  v_0 \tan\xi -2 \lambda  v_0 }{1+u_0 v_0}, \nonumber\\
&\qquad \qquad 
 \frac{  (v_0-u_0) \sec\xi -\eta\left(1+v_0^2\right) \tan\xi
 -\lambda  \left(1+v_0^2\right) }{1+u_0 v_0}, 
 \tan (\xi )
    \bigg)\,.
\end{align}

By the same argument as before, new bulk caustics only occur due to the identifications required to take the quotient to obtain BTZ from AdS$_3$. This rules out the possibility that the caustic cuts off the inward pointing light ray before it hits the singularity, since the singularity is reached within a fundamental domain of the identification. Instead the caustics will extend from the singularity back to the boundary where the generators on opposite sides of the radial light ray meet at the surface fixed by the identification.  
This time the last component of $L_1$ is odd under $\xi \rightarrow -\xi$, so that the caustic occurs on the identified planes $P_{plane}(\pi R)$ and $P_{plane}(-\pi R)$.

\begin{figure}
	\begin{subfigure}[b]{0.45\textwidth}
	\centering
		\includegraphics[width=0.9\textwidth]{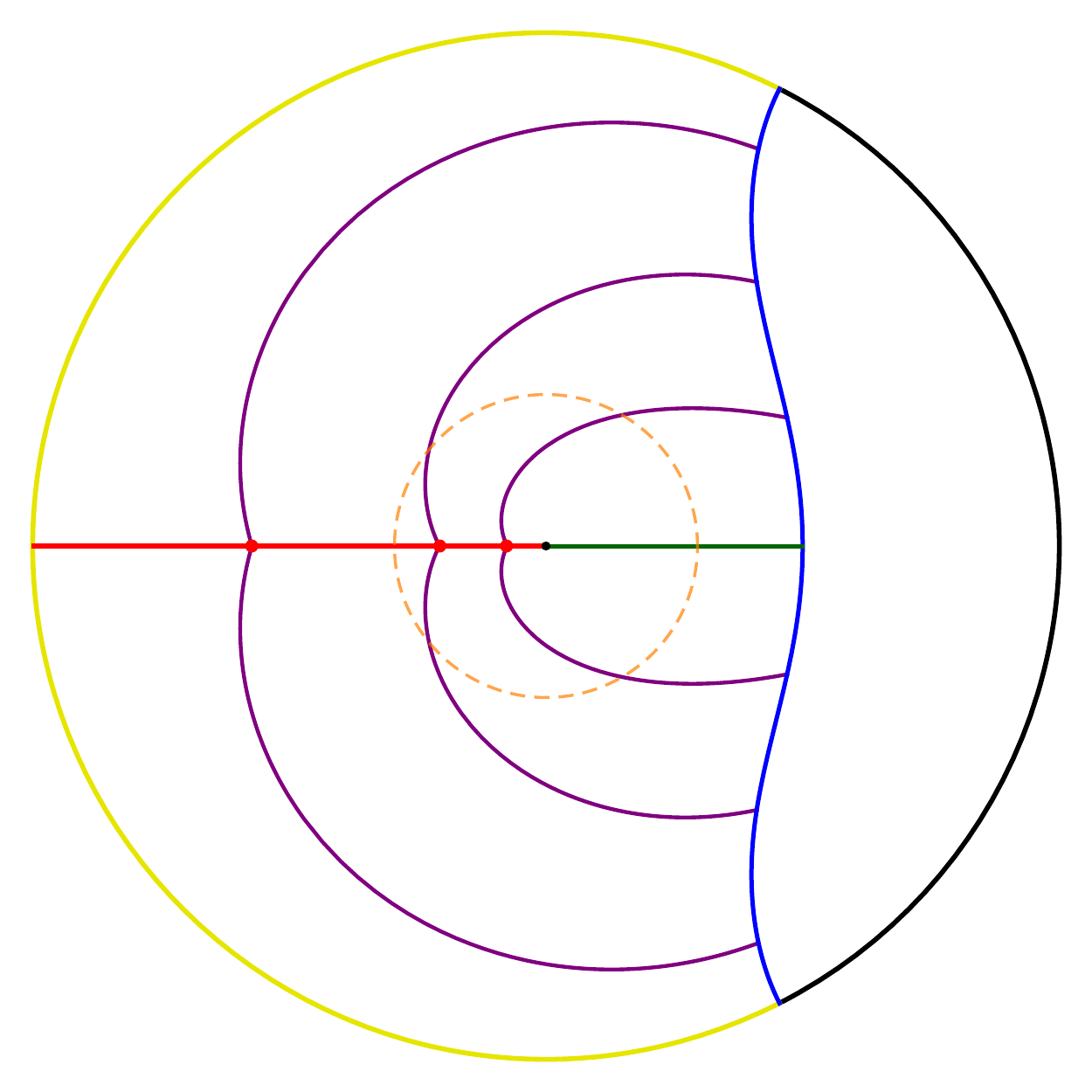}
		\caption{Schwarzschild-like $(r,\theta)$ diagram}
		\vskip 2mm
		\label{fig: caustic btz t projected out}
	\end{subfigure}
	\begin{subfigure}[b]{0.45\textwidth}
	\centering
	\includegraphics[width=.9\textwidth]{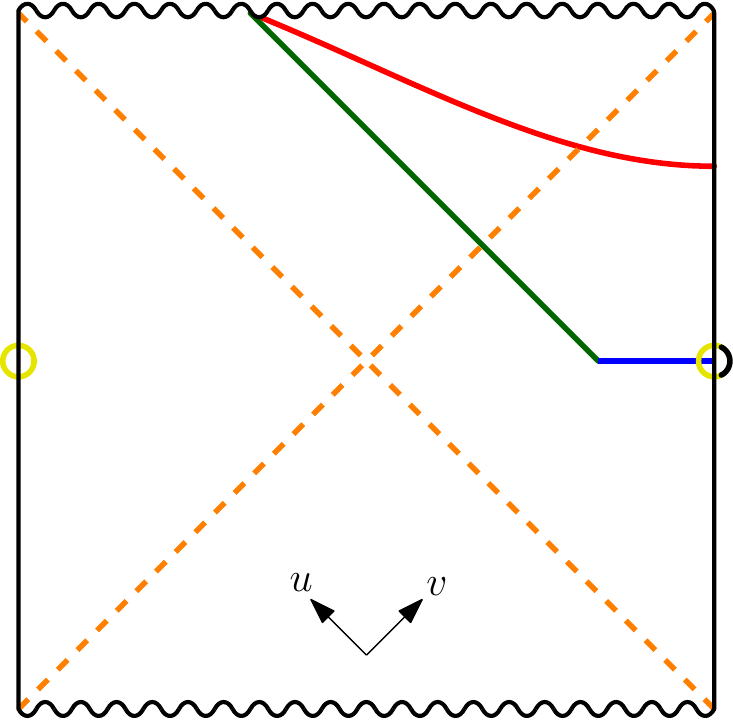}
		\caption{Penrose diagram}
		\vskip 2mm
		\label{fig: caustic btz penrose}
	\end{subfigure}
	\caption{The red line represents the future caustic of a boundary region in the one- and two-sided BTZ black hole. The boundary region, shown in yellow, comprises more than half of the $t=0$ slice of the right boundary and the complete left boundary time slice. The HRT surface is shown in blue and is chosen at $r_0 = 1$. A few representative orthogonal light rays are drawn in purple in (a) and meet at the caustic. The radial light ray reaching the singularity is shown in green. The horizon is chosen at $R=0.5$ and is indicated in dashed orange.}
	    \label{fig: caustic penrose btz}
\end{figure}

The solutions to $L_1(\lambda_1,\xi)\cdot P_{plane}(\pi R)=0$ and $L_1(\lambda_2,-\xi)\cdot P_{plane}(-\pi R)=0$ are 
\begin{align}
\lambda_1 = - \frac{Y(\xi)\cdot P_{plane}(\pi R) }{N_1\cdot P_{plane}(\pi R) } \,, \qquad
\lambda_2 = - \frac{Y(-\xi)\cdot P_{plane}(-\pi R) }{N_1\cdot P_{plane}(-\pi R) } \,.
\end{align}
The caustic is located at 
\begin{align}
L_1(\lambda_1,\xi) &= \bigg(
\frac{\left(v_0^2+1\right) \sec\xi 
+\left(v_0^2-1\right) \coth \pi  R \tan\xi }{2 v_0},
\coth \pi R \tan\xi , \nonumber\\ &\qquad \qquad
\frac{\left(v_0^2-1\right) \sec\xi +\left(v_0^2+1\right) \coth \pi  R \tan\xi }{2 v_0},
\tan \xi  \bigg) \,, 
 \\
L_1(\lambda_2,-\xi) &= \bigg(
\frac{\left(v_0^2+1\right) \sec\xi 
+\left(v_0^2-1\right) \coth \pi  R \tan\xi }{2 v_0},
\coth \pi R \tan\xi , \nonumber\\ &\qquad \qquad
\frac{\left(v_0^2-1\right) \sec\xi +\left(v_0^2+1\right) \coth \pi  R \tan\xi }{2 v_0},
-\tan \xi  \bigg) \,. \nonumber
\end{align}
In the hyperpolar coordinates of \eqref{eqn:hyperpolar_BTZ}, $L_1(\lambda_1,\xi)$ and $L_1(\lambda_2,-\xi)$ are located at $\mu = \pm \pi R$ respectively, which are to be identified, confirming that this is the location of a caustic. This can be related to a position in the Kruskal-like coordinates by inverting \eqref{eqn:BTZcoords}.
This determines the future caustic to lie along $\theta=\pm \pi$ at
\begin{align}
v(u) = v_0 \frac{u v_0 + \cosh \pi  R}{1 + u v_0 \cosh \pi  R}\,.
\label{eqn:2-sided_BTZ_caustic}
\end{align}
This is illustrated in figure \ref{fig: caustic btz penrose}.
Notice that \eqref{eqn:2-sided_BTZ_caustic} is independent of both the boost of the boundary region, $\eta$, and of $u_0$. 
Similarly to the result we found in the conical deficit, the caustic only depends on the location of the future tip of the boundary causal diamond.  By following the caustic out to the boundary, that is comparing the light ray in the direction of
\begin{align}
\lim_{\xi\rightarrow \frac\pi2 } \frac{\sinh\pi R}{\tan\xi} L_1(\lambda_1,\xi) &=
\bigg(
\frac{\left(v_0^2+1\right)  \sinh \pi  R 
+\left(v_0^2-1\right) \cosh \pi  R  }{2 v_0},
\cosh \pi R , 
\\ &\qquad \qquad
\frac{\left(v_0^2+1\right) \cosh \pi R + \left(v_0^2-1\right) \sinh \pi  R }{2 v_0},
\sinh \pi R  \bigg) \,, \nonumber
\end{align}
to our parametrisation of the boundary, \eqref{eqn:BTZ_bdy}, we see that this future tip is located at $\theta=\pm \pi$ and
\begin{align}
t(r=\infty) = \pi + \frac{1}{R} \log v_0\,.
\end{align}

\subsection{Localisability in two-sided BTZ}
In this section we will discuss the localisable region in the two-sided eternal BTZ black hole. This region will be quite different from that in the one-sided BTZ black hole that could be formed by collapse, due to the existence of spacelike geodesics stretching from one boundary to the other. In the two-sided case, the entanglement shadow is behind the horizon near the singularity. In fact, the entire spacetime is probed by spacelike geodesics stretching between the two boundaries, but the length of these geodesics grows as they approach the singularity. Since regions to which these geodesics can be anchored also admit candidate extremal surfaces consisting of disconnected geodesics that stay outside of the horizon, these disconnected geodesics will dominate once the geodesic that crosses gets close enough to the singularity. This leads to an entanglement shadow near the singularity in the interior of the black hole \cite{shadows14}.

This entanglement shadow behind the horizon allows us to use our lemma \ref{thm:need-HRT} to argue that there is a non-localisable region near the horizon. 
In particular, given the explicit form of the entanglement wedges of regions that include the entire left boundary as well as a subregion of the right boundary derived in the previous section (see figure \ref{fig: caustic btz penrose}), we confirm that everything to the left, on the conformal diagram, of the central ingoing light ray that hits the singularity is included in the entanglement wedge. This confirms the picture in figure 5 used by \cite{SW} in their argument establishing the non-localisability of a region near the singularity of the two-sided BTZ.

\subsection{Localisability in one-sided BTZ}
\begin{figure}
\centering
		\includegraphics[width=.6\textwidth]{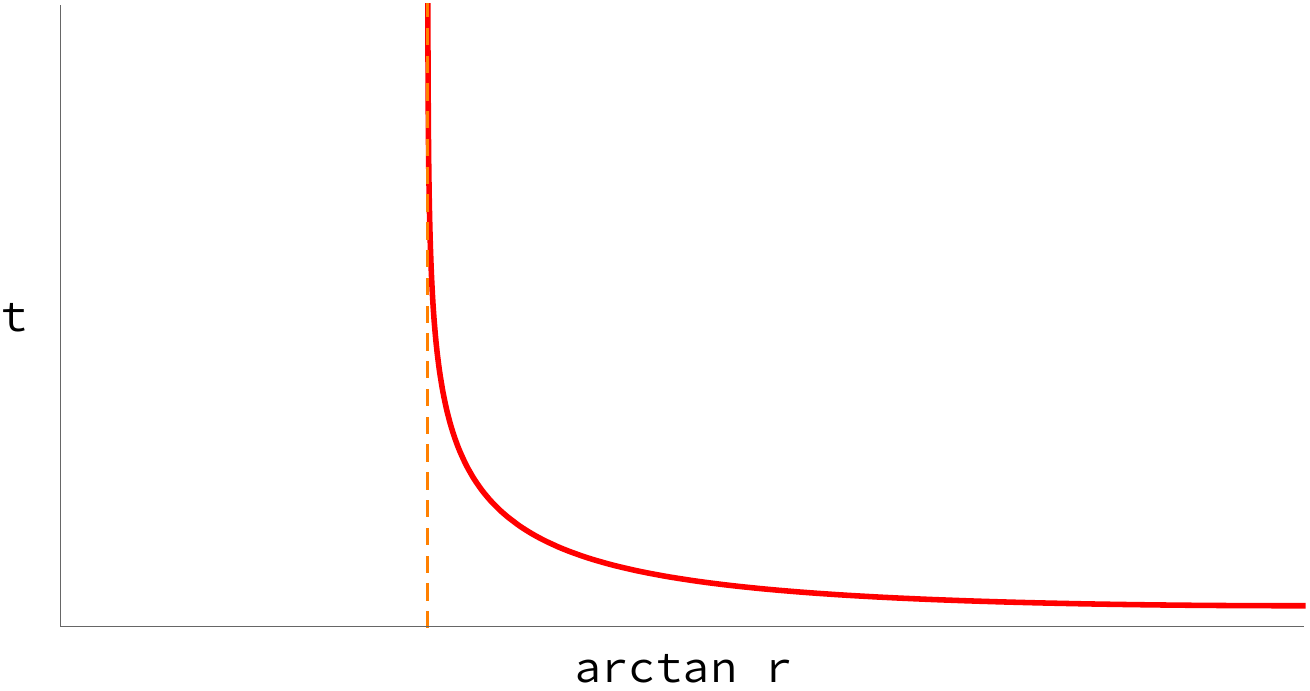}
		\vskip 2mm
		\caption{The time dependence of the caustic outside the horizon of a BTZ black hole as a function of $r$, with a horizon of radius $R = 0.5$ for an HRT surface with $r_0=1$. The solid red line is the caustic and the orange dashed line depicts the location of the horizon.}
    \label{fig: caustic t(r) btz}
\end{figure}

If we only have access to one boundary of the BTZ black hole, then there is a region near the horizon that cannot be reached by HRT surfaces, much as in the conical deficit spacetime \cite{plateaux13,shadows14}. Here again we could try to use the strategy proposed by \cite{SW} for the conical deficit to localise points in this region. However, from figure \ref{fig: caustic t(r) btz} we can see that this strategy will not work for the same reasons that it failed for $0<\alpha\leq \frac12$ in the conical deficit. To analyse this, it is useful to introduce Schwarzschild-like coordinates covering the exterior region of BTZ. These have the form
\begin{align}
X_{BTZ'}^A &= \left( \frac{\sqrt{r^2-R^2}}{R} \sinh R t, \frac{r}{R} \cosh R \theta,  
\frac{\sqrt{r^2-R^2}}{R} \cosh R t, \frac{r}{R} \sinh R \theta \right) \,, \\
ds^2 &= dX_{BTZ'} \cdot dX_{BTZ'} = -(r^2-R^2) dt^2 + \frac{dr^2}{r^2-R^2} +r^2 d\theta^2\,.
\end{align}
These coordinates are related to the Kruskal-like ones by\footnote{The Schwarzschild-like coordinates cover the right exterior region, where $v>0$ and $u<0$.}
\begin{align}
\frac{r}{R}=& \frac{1-uv}{1+uv}\,, & v =& \sqrt{\frac{r-R}{r+R}} e^{R t}  \,, \\
e^{R t} =& \sqrt{\frac{v}{-u}}\,, & u=&-\sqrt{\frac{r-R}{r+R}} e^{-R t} \,.
\end{align}
An example of a caustic in BTZ is shown in figure \ref{fig: caustic penrose btz} in both set of coordinates.

In the previous section, our HRT surfaces were parametrised by $(u_0,v_0)$. We can use the time-translation symmetry of the metric in Schwarzschild-like coordinates to fix $t_0=0$. This means that $u_0 = -v_0$ and 
\begin{align}
r_0 = R \frac{1+v_0^2}{1-v_0^2} \,.
\end{align}

Applying this change of coordinates to the expression for the caustics obtained in \eqref{eqn:2-sided_BTZ_caustic}, the caustic is found to lie at 
\begin{align}
\label{eq: caustic btz}
t(r) &= 
\frac{1}{R} \left( 
\mathrm{arctanh}\, \frac{ \sqrt{R^2 + r^2 \sinh^2 \pi R} }{r \cosh \pi R} - \mathrm{arctanh}\, \frac{R}{r_0}
\right)\,,\\
\partial_r t(r) &= - \frac{ R \cosh \pi R }{(r^2-R^2) \sqrt{R^2+r^2 \sinh^2 \pi R} }\,.
\end{align}
This expression makes explicit that $t(r)$ is a monotonically decreasing function of $r>R$ which diverges as $r \rightarrow R$, as shown in figure \ref{fig: caustic t(r) btz}. Note that \eqref{eq: caustic btz} can also be found from \eqref{eq: caustic conical} by analytically continuing $\alpha \rightarrow i R$. 

This implies that any entanglement wedge whose caustic passes through the point $(r_*,t_*,\theta_*)$, will include all the points along a line at fixed $t$ going inwards from this point, that is the points \begin{align}
(r,t_*,\theta_*) ~ \mathrm{for}~ r \in (R,r_*) \,.
\end{align}
By the same logic used in the conical deficit spacetime, this demonstrates that the non-localisable region of the one-sided BTZ black hole coincides with the entanglement shadow.

\section{Outlook}
In this work we studied the detailed form of the caustics bounding the entanglement wedges in simple spacetimes. Entanglement wedges play an essential role in understanding the emergence of bulk locality \cite{HQECC14,JLMS15,EW_reconstruction16} and in the diagnosis of bulk locality from the error correcting structure of holography proposed in \cite{SW}, the shape of the caustics bounding these entanglement wedges is important in determining the bulk region for which local bulk operators can be identified as local using boundary techniques. 

Our analysis of the detailed form of these caustics revealed unexpected features that contradicts the assumptions in some of their analysis, while confirming those made in other parts. In particular, in the setting of asymptotically AdS$_3$ spacetimes, we find a non-localisable region near the horizon of a one-sided BTZ black hole and near conical singularities with sufficiently large angular deficits which coincides with the entanglement shadow. In the conical deficit, the non-localisable region appears when the caustics bend sufficiently sharply away from the trajectory of the light rays approaching the conical singularity leading to a sharp corner in the entanglement wedge near the conical singularity.

It would be interesting to better understand the caustics appearing on the boundaries of entanglement wedges in higher dimensions. Since the lightsheets will be higher dimensional objects, there is a more complicated zoo of caustics that could occur with the possibility of lower dimensional caustics where higher dimensional caustics pinch off. There is also a variety of boundary regions that can be considered, whereas a 2-dimensional boundary only admits intervals. Less symmetric boundary regions will generally lead to the presence of more caustics, even in empty AdS space. A better understanding of the possible shapes of these caustics is required to understand the entanglement wedges in these geometries with all the ensuing implications for understanding bulk locality.

The study of the caustics in more complicated settings, such as higher dimensions will require numerical techniques. Finding the locations of caustics becomes the problem of finding where light rays cross in the bulk. As the number of dimensions grows the number of parameters which must be tuned for this to occur grows as well, not to mention that even finding the HRT surfaces, which emit these lightsheets, in higher dimensions requires solving PDEs rather than ODEs. In appendix \ref{sec: Numerical approach to the lightsheet construction}, we discuss a numerical approach to determining the caustics in the simple setting we studied in this work. This numerical approach was used to confirm our analytic results and provides a starting point for further studies in more complicated settings.

\section*{Acknowledgements}
We would like to thank Ben Craps and Sean Weinberg for discussions.

This work is supported in part by FWO-Vlaanderen through projects G044016N and G006918N and by Vrije Universiteit Brussel through the Strategic Research Program ``High-Energy Physics.'' M.~D.~C.~is supported by a PhD fellowship from the Research Foundation Flanders (FWO).
C.~R.~also acknowledges support from the Natural Sciences and Engineering Research Council of Canada (NSERC) funding reference number PDF-517316-2018 and from a Postdoctoral Fellowship from the Research Foundation Flanders (FWO).

\appendix
\section*{Appendix}
\section{Numerical approach to the lightsheet construction}
\label{sec: Numerical approach to the lightsheet construction}
We demonstrate the numerical approach to the (future) lightsheet construction of the entanglement wedge for boundary regions comprising more than half of a spatial slice (possibly not a constant time slice) of a conical deficit spacetime, based on \cite{HRT07}. 

We are given a spatial geodesic anchored on a boundary region in a conical deficit. The lightsheet starting from the geodesic and reaching the boundary diamond associated to the boundary region can be found by computing light rays orthogonal to the HRT surface and pointing towards the boundary region of interest. The fact that light rays are null leads to a constraint in the form of a differential equation obtained by setting the line element to \eqref{linecon} to zero. Using an affine parameter $\lambda$, this is
\begin{equation}
\label{odeCondefLight}
    0=-\left( \alpha^{2}+r(\lambda)^2\right)\left(\frac{dt(\lambda)}{d\lambda}\right)^2
    + \frac{1}{\alpha^{2} +r(\lambda)^2}\left(\frac{dr(\lambda)}{d\lambda}\right)^2
    +r(\lambda)^2 \left(\frac{d\theta(\lambda)}{d\lambda}\right)^2.
\end{equation}
This can be turned into a first order ordinary differential equation by using conserved quantities associated to Killing vectors of the spacetime. The metric \eqref{linecon} depends neither on $t$ nor on $\theta$, which leads to two Killing vectors $\partial_t=\delta_t^\mu\partial_\mu$ and $\partial_\theta=\delta_\theta^\mu\partial_\mu$, and their associated conserved quantities
\begin{align}
\label{EConDef}
    E=&-g_{\rho\mu}\delta_t^\mu\frac{dx^\rho(\lambda)}{d\lambda}=-g_{tt}\frac{dt(\lambda)}{d\lambda}
    =\left(\alpha^{2}+r(\lambda)^2\right)\frac{dt(\lambda)}{d\lambda}
    \\
    &\text{and}\notag
    \\
    \label{PConDef}
    P_{\theta}=&-g_{\rho\mu}\delta_\theta^\mu\frac{dx^\rho(\lambda)}{d\lambda}=-g_{\theta\theta}\frac{d\theta(\lambda)}{d\lambda}
    =-r(\lambda)^2\frac{d\theta(\lambda)}{d\lambda}.
\end{align}
The equation describing lightlike geodesics in a conical deficit therefore becomes
\begin{align}
    \left(\frac{dr(\lambda)}{d\lambda}\right)^2
    =E^2
    -\frac{P_\theta^2}{r(\lambda)^2}\left( \alpha^{2}+r(\lambda)^2\right).
        \label{odeCondefLightEP}
\end{align}
The ratio between $E$ and $P_\theta$ is fixed up to a sign by the demand that the light ray be orthogonal to the HRT surface given by ($r_{ex}(\theta),t_{ex}(\theta)$). To see this one realises that since the surface has two dimensions less than the spacetime it is fixed by two constraints. Namely
\begin{equation}
    \varphi_1(x^\mu)=t_{ex}(\theta)-t=0\label{ConsConDef1}
\end{equation}
and
\begin{equation}
    \varphi_2(x^\mu)=r_{ex}(\theta)-r=0\label{ConsConDef2}.
\end{equation}
Any vector orthogonal to the surface must be a linear combination of the covariant derivatives of the two constraints $\nabla^\nu\varphi_1$ and $\nabla^\nu\varphi_2$. The components with lowered indices of a generic such vector, $N$, can thus be written as
\begin{equation}
\label{orthformgen}
    N_\nu= \nabla_\nu\varphi_1+\mu_{\pm}\nabla_\nu\varphi_2,
\end{equation}
With the constraints \eqref{ConsConDef1} and \eqref{ConsConDef2} this becomes
\begin{equation}
\label{orthformConDef}
    N_\nu= \delta^\theta_\nu t_{ex}'(\theta_0)-\delta_\nu^t+\mu_{\pm}(\delta_\nu^\theta r_{ex}'(\theta_0)-\delta_\nu^r),
\end{equation}
where the angle $\theta_0$ indicates on which point on the HRT surface the light ray starts.

Saying the light ray parametrised by $x^\mu(\lambda)$ is orthogonal to the HRT surface means that the components of the tangent vector, $\frac{dx^\mu(\lambda)}{d\lambda}$, are of this form, i.e. $N^\mu=\frac{dx^\mu(\lambda)}{d\lambda}$. One thus obtains the ratio between $P_\theta$ and $E$,
\begin{equation}
\label{PErat}
    \frac{P_\theta}{E}=\frac{-g_{\theta\theta}\frac{d\theta(\lambda)}{d\lambda}}{-g_{tt}\frac{dt(\lambda)}{d\lambda}}\Bigg|_{\lambda=\lambda_0}
    =-\mu_\pm r_{ex}'(\theta_0)-t_{ex}'(\theta_0).
\end{equation}
 To determine $\mu_\pm$ one can exploit the null norm of \eqref{orthformConDef}
\begin{align}
    0=&g^{\mu\nu}N_\mu N_\nu
    =g^{tt}+\mu_\pm^2g^{rr}+ \left(t_{ex}'(\theta_0) + \mu_\pm r_{ex}'(\theta_0)\right)^2 g^{\theta\theta}
    \end{align}
such that
\begin{align}
    \mu_\pm=& \frac{-g^{\theta \theta}r_{ex}'(\theta_0)t_{ex}'(\theta_0) \pm \sqrt{-g^{tt}(g^{rr}+r_{ex}'(\theta_0)^2 g^{\theta \theta}) - g^{rr}g^{\theta \theta} t_{ex}'(\theta_0)^2}}{g^{rr}+ r_{ex}'(\theta_0)^2g^{\theta\theta}}.
 \label{mupm}
\end{align}
The two roots reflect the possibility to have a lightsheet that points to either of the two boundary regions associated to the geodesic. The largest boundary region corresponds to the choice of $\mu_+$. The light rays $(r(\lambda),\theta(\lambda),t(\lambda))$ forming the lightsheet $(r(\lambda,\theta_0),\theta(\lambda,\theta_0),t(\lambda,\theta_0))$ can then be solved for numerically by setting $E=1$, as this amounts to a choice of the affine parameter for the light rays, and solving \eqref{odeCondefLightEP} with the following conditions:
\begin{itemize}
\item $r(0)=r_{ex}(\theta_0)$
\item $t(0)=t_{ex}(\theta_0)$
\item $\theta(0)= \theta_0$
\item $t'(\lambda)= -g_{tt}^{-1}$
\item $\theta'(\lambda)= r^{-2}(\lambda)\left(\mu_+ r'_{ex}(\theta_0)+t'_{ex}(\theta_0)\right)$.
\end{itemize}

These light rays determine the lightsheet, except that they must be terminated when they cross. This is an equation of the form
\begin{align}
(r(\lambda_1,\theta_1),\theta(\lambda_1,\theta_1),t(\lambda_1,\theta_1)) = (r(\lambda_2,\theta_2),\theta(\lambda_2,\theta_2),t(\lambda_2,\theta_2)) \,. 
\end{align}
This can be solved numerically by fixing the generator of interest by fixing $\theta_1$ and sweeping through the other parameters, namely the other generator with which it intersects, $\theta_2$, and where along these generators the intersection occurs, $(\lambda_1,\lambda_2)$. For these 2-dimensional lightsheets, the caustics will be localised along a 1-parameter family of points.

\providecommand{\href}[2]{#2}\begingroup\raggedright\endgroup

\end{document}